\documentclass[a4paper,fleqn,usenatbib]{mnras}

\usepackage[T1]{fontenc}
\usepackage{ae,aecompl}

\usepackage{graphicx}	
\usepackage{amsmath}	
\usepackage{amssymb}	
\usepackage{multirow}
\usepackage{physics}

\newcommand{\hmrad}{r_{1/2,\,*}}
\newcommand{\msun}{\mathrm{M}_\odot}
\newcommand{\jzc}{j_z/j_c}
\newcommand{\jpc}{j_p/j_c}
\newcommand{\e}{e/|e|_\mathrm{max}}
\newcommand{\led}{\lambda_\text{Edd}}

\title[Polar-rings in the TNG50]{Polar-ring galaxies in the Illustris TNG50 simulation}

\author[D. V. Smirnov, A. V. Mosenkov, V. P. Reshetnikov]{
	Daniil V. Smirnov$^{1}$ \thanks{E-mail: smirnovdv99@gmail.com}, Aleksandr V. Mosenkov$^{2}$, Vladimir P. Reshetnikov$^{1}$
	\\
	$^{1}$St.Petersburg State University, 7/9 Universitetskaya nab., St.Petersburg, 
	199034 Russia\\
	$^{2}$Department of Physics and Astronomy, N283 ESC, Brigham Young University, Provo, UT 84602, USA 
}

\date{Accepted 2022. Received 2022; in original form 2022}

\pubyear{2023}

\begin{document}
	\label{firstpage}
	\pagerange{\pageref{firstpage}--\pageref{lastpage}}
	\maketitle
	
	\begin{abstract}
		
		Polar-ring galaxies (PRGs) are an outstanding example of galaxies with misaligned kinematics where a typically red central galaxy is surrounded by a large-scale ring or disk of stars, gas and dust oriented almost perpendicular to the main body. It is believed that polar structures are formed in a secondary event after the assembly of a central galaxy, but due to their scarcity, their formation paths are not well constrained yet. We present a study of PRGs from TNG50 cosmological simulations, focusing on the origin of their polar structures. Based on the synthetic images and baryonic mass distribution, we found 6 galaxies with stellar polar rings. Using Supplementary Data Catalogues and available particle data, we confirm that the selected galaxies are direct analogs of real PRGs. In our sample, the polar structures are a result of the close interaction between the host galaxy and its companion. We track two formation paths for the stellar polar rings in our sample: (i) star formation in the accreted gas, (ii) tidal disruption of the satellite’s stellar component. Rings formed during the first scenario are, on average, bluer and younger than ones formed due to the satellite disruption. We report a steady increase of the ring's inclination around the two most massive galaxies across a few billion years with a rate of $\approx8^\circ/$Gyr. The formation of a polar structure in some cases can increase the nuclear activity of the central galaxy and/or turn the active nucleus off completely.
		
	\end{abstract}
	
	
	
	
	\section{Introduction}
	
	Polar-ring galaxies (PRGs) are a unique type of extragalactic objects which consist of two large-scale subsystems -- a central (host) galaxy and an extended ring or a disk of stars and gas oriented almost orthogonally to the galactic plane. In most cases, two components are significantly different in appearance: host galaxies (HGs) are usually gas-free early-type galaxies (ETGs) while polar structures (PS) with blue colours and signs of active star-formation resemble late-type galaxies \citep{whit1990, Finkelman2012, Reshetnikov2015}. Polar rings are kinematically decoupled from the main body of the central galaxy showing an ordered rotation around its major axis (e.g. \citealt{em2019}).
	
	The existence of the two distinct subsystems in PRGs requires a specific explanation within the assembly history of the galaxy. Through the years a number of different mechanisms were proposed to explain the structure of PRGs. Modern formation scenarios assume that polar structures emerged as a result of a separate event after the central galaxy was formed. As such secondary event, \citet{bekki1997, bekki1998} suggested a head-on collision of a future host galaxy with an orthogonally oriented disk galaxy (merger scenario). They showed that such an interaction of two late-type spirals with a sufficiently low relative velocity could result in the formation of a realistic and stable polar ring. In a later study \citet{bc2003} investigated the collision of a flattened early-type galaxy (E10) with a gas-rich spiral. They confirmed the sensitivity of the polar ring formation to the relative velocity of galaxies and found that a sufficiently small impact parameter is required.
	
	The accretion scenario \citep{Schweizer1983, rs1997} provides another explanation for the formation of polar structures. In this variant, the ring is formed due to a close interaction of two galaxies (a future host and a donor) accompanied by a tidal accretion. The donor is presumed to move along a near-polar orbit and possess large amounts of gas. Numerical studies showed that this scenario is less sensitive to the distance between galaxies and relative velocity, as well as the internal composition of the future host \citep{bc2003}. The scenario proposed in \citet{Rix1991} and \citet{Katz1992} with the capture into a near-polar orbit and the destruction of a gas-rich satellite can be considered an extreme case of the accretion scenario.
	
	Cold filament accretion was also employed to explain the formation of polar rings. It is widely assumed that galaxies, especially in the early stages of their evolution, accrete a large amounts of cold gas from the cosmic filaments and intergalactic medium (IGM). Accounting for such accretion within the framework of hierarchical assembly in the $\Lambda$CDM cosmology explains important observational data regarding the formation of galaxies \citep{Dekel2009, Agertz2009}. It was found that the formation of some ringed galaxies may be a result of cold filament accretion (e.g. \citealt{Stanonik2009, Silchenko2023}). If the angular momentum of the accretting gas lies closely to the disk plane of the central galaxy, then a polar structure may be formed as a result. Using cosmological zoom-in simulations, \citet{maccio2006} and \citet{brook2008} showed that a lifelike polar-ring galaxy could form due to cold filamentary accretion.
	
	Although the question of the PRGs' origin arose shortly after their discovery, there is still no clear answer to it, as well as a reliable observational test which allows to distinguish the formation mechanism in each case. Studies utilising indirect s indicate that all of the mentioned scenarios may occur in real life. Thus, while studying deep images of 13 PRG candidates, \citet{Mosenkov2022} found that most of the objects demonstrate low surface brightness (LSB) features, typically associated with recent interaction. Several galaxies exhibit bright spots embedded in the LSB structures associated with the nuclei of a disrupted victim galaxy while others do not, favouring the accretion scenario. \citet{em2019} analysed long-slit spectroscopic observations of 15 polar rings. They found that the rings' gas has a sub-solar metallicity which has a flat distribution and does not depend on the host's luminosity. These findings clearly indicate that the gas in the rings is of external origin. Authors rule out the cold accretion scenario as the derived metallicities are much higher than expected and are more consistent with the merging or accretion scenario. On the other hand, an extensive spectrophotometric study of NGC4650A  -- a classical example of polar-ring galaxy -- revealed low metallicity values in the ring, the absence of metallicity gradient and signs of a continuous infall of metal-poor gas which favours cold accretion scenario for this galaxy \citep{Iodice2006, Spavone2010}.
	
	The currently existing scenarios of the PRG formation have been repeatedly tested in idealised simulations \citep{Rix1991, rs1997, bekki1998, bc2003}. This approach makes it possible to follow the process of the formation and evolution of polar-rings for a predetermined scenario and a set of initial conditions in detail. However, it is almost impossible to study the prevalence, relative frequency, and realism of the implied initial conditions with such a modelling \citep{bc2003}. In contrast, the use of zoom-in simulations makes it possible to take into account cosmologically motivated starting conditions by considering an initially large volume of space, from which an area is subsequently extracted for detailed modelling. However, the end result of such simulations is still one single galaxy with the desired morphology and its close environment, which does not allow for a study of the formation mechanisms statistics on cosmological scale. Modern cosmological simulations with sufficiently high spatial resolution, such as EAGLE \citep{EAGLE}, IllustrisTNG \citep{Pillepich2019, Nelson2019}, SIMBA \citep{SIMBA}, Horizon-AGN \citep{Horizon} can help to overcome this obstacle. Due to their large volume, these simulations cover a wide range of spatial scales which makes it possible to study the formation and evolution of galaxies in the cosmologically motivated environment, determined by the evolution of the Universe. Studying the formation and evolution of PRGs in high-resolution cosmological simulations can help us better understand the origin of these rare objects, estimate the frequency of different formation scenarios, and derive possible observable indicators of the formation mechanisms.
	
	In this paper we aim to explore the formation paths of galaxies with stellar polar structures at $z=0$ with the help of cosmological simulations TNG50. A total of 6 galaxies with apparent morphology similar to the real PRGs were found in the simulation box. It is shown that in terms of characteristics these objects are indeed analogs of observed PRGs. Using merger trees we study what processes lead to the formation of polar structures and explore their impact on the galaxy.
	
	This paper is organised as follows. In section~\ref{sec:search} we describe the sample of PRGs we selected from all galaxies in simulation and the procedure we used to separate polar structures from the central galaxies. Section~\ref{sec:characteristics} compares the main characteristics of simulated galaxies and the real ones. In section~\ref{sec:formation} we explore the formation paths of polar structures in simulations and study the evolution of their key parameters. Section~\ref{sec:conclusions} contains discussion of the acquired results and conclusions of this manuscript.
	
	\section{Search for PRGs in TNG50}\label{sec:search}
	
	\subsection{TNG50-1 simulations}
	
	IllustrisTNG\footnote{\url{https://www.tng-project.org/}} \citep{Nelson2018, Springel2018, Pillepich2018, Naiman2018, Marinacci2018, Pillepich2019, Nelson2019} is a suite of cosmological magnetohydrodynamical simulations that track the formation and evolution of galaxies in a $\Lambda$CDM universe. The simulations utilise moving-mesh code \textsc{AREPO} \citep{AREPO} and take into account such important astrophysical processes as gas heating and cooling in a variable UV background, star formation and stellar evolution, SN feedback and stellar winds, ISM enrichment, growth of SMBH and their feedback (see \citealt{TNG_Pillepich, TNG_Weinberger} for full details on the TNG model). The free parameters of the model were calibrated to match the observed scaling relations of galaxies at $z=0$. The cosmological parameters were set in accordance with the results of the \citet{Planck2015}.
	
	A Friends-of-Friend \citep{Fof} and \textsc{Subfind} \citep{Subfind} algorithms were used to identify gravitationally bound structures in the dark matter distribution and baryonic subhalos (galaxies) inhabiting them. Subhalos are tracked between snapshots via \textsc{Sublink} merger trees, which allows one to easily follow the evolution of individual galaxies through cosmic time.
	
	Results of IllustrisTNG have successfully proved useful for the study of morphologically peculiar galaxies and especially their formation mechanisms in cosmological context (e.g., \citealt{Zhu2018, Yun2019, Semczuk2020, Sales2020, Lokas2021, Lokas2021b, Khoperskov2021, Zee2022, Perez2022, Lokas2022}). This study utilises the results of TNG50(-1) -- the smallest but the most resolved of the available IllustrisTNG simulations. It has a box size of 51.7\,cMpc and contains $2\times 2160^3$ total initial resolution elements, which provides a baryonic mass resolution of $8.5\times 10^4\,\msun$ and an average gas cell size around 70-140\,pc. Such parameters make TNG50 a perfect simulation for the task: due to its high mass and spatial resolution comparable to zoom-in simulations, potential polar structures will have enough particles to adequately study them, while a relatively large box size produces a representative sample of galaxies across different environments.
	
	\subsection{Galaxy selection}
	
	Various approaches can be used to search for galaxies with peculiar morphology in cosmological simulations; for example, selecting galaxies based on their kinematic properties \citep{Khoperskov2021}, structural parameters \citep{Lokas2021b, Lokas2022}, visual morphology of their stars and gas distribution \citep{Semczuk2020} or synthetic images \citep{EAGLE_rings}. Unfortunately, the high variety of observed polar-ring galaxies does not allow for a robust selection criteria, both in observations and simulations. It is possible to distinguish PRGs from ``normal" galaxies based on the distribution of angles between the angular momentum vector of the stellar particles and the galaxy's plane or with the help of full-fledged kinematic decomposition (see section~\ref{subsec:kin_decomp}). These techniques are computationally expensive and still require visual control afterwards, so in this study we focus on a preliminary sample of polar-ring galaxies selected based on their synthetic images.
	
\begin{figure*}
		\centering
		\includegraphics[width=\textwidth]{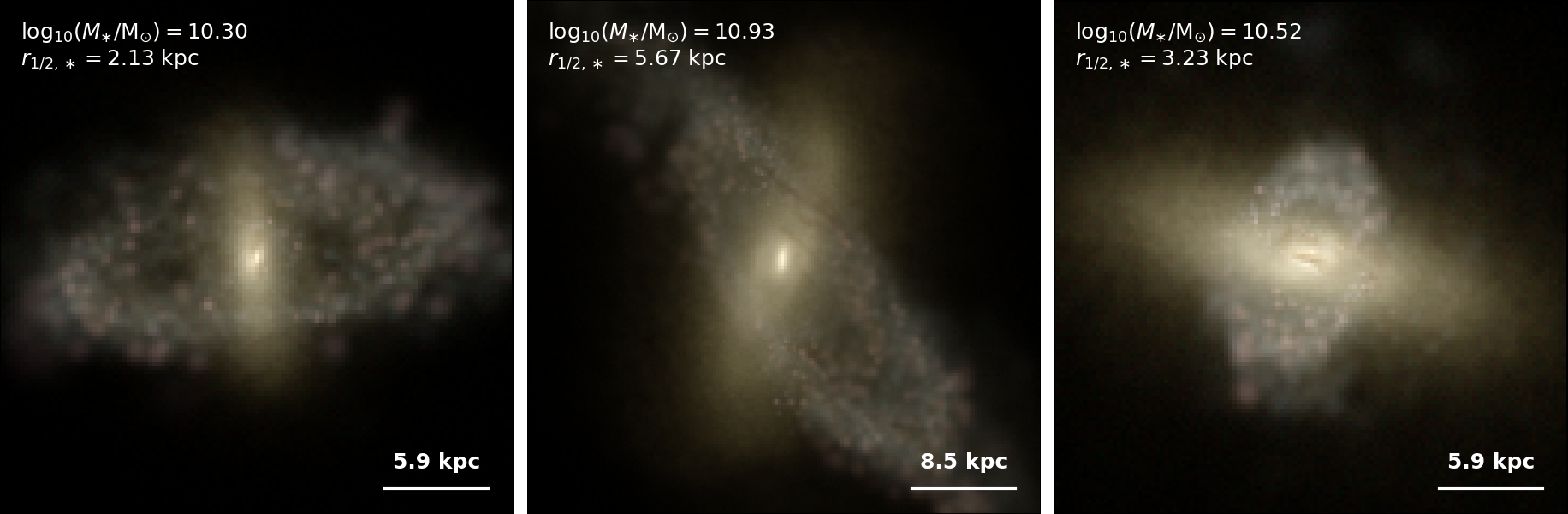}
		\caption{Examples of synthetic images from IllustrisTNG Infinite Gallery of PRG candidates generated with SKIRT following \citet{Rodriguez-Gomez2019}. Bars on the bottom right indicate the scale of the images and text labels on the upper left indicate the total stellar mass and the stellar half-mass radius of each galaxy.}
		\label{fig:synt_im}
	\end{figure*}

	To search for PRGs, we have utilised the TNG50 Infinite Gallery page\footnote{\url{https://www.tng-project.org/explore/gallery/rodriguezgomez19b/}}, which contains synthetic images of over 1600 galaxies from TNG50-1 obtained at $z=0.05$ (snapshot 95). The images were generated with the SKIRT radiative transfer code \citep{SKIRT1, SKIRT2} in the $g,\, r,\, i$ Pan-STARRS filters (a detailed description of the technique originally applied to galaxies from TNG100 is given in \citealt{Rodriguez-Gomez2019}). After looking through all the available images, we have selected 31 candidates with an apparent morphology similar to that of real PRGs. We searched for galaxies with two nearly orthogonal flat subsystems (``main" disk and a polar structure) with coinciding centres. We only selected galaxies where the PS is closed in order to exclude transient tidal structures from our sample. \autoref{fig:synt_im} shows synthetic images of three of these galaxies. The described method is obviously subject to selection effects (galaxies are oriented randomly on the images, and not all sufficiently massive galaxies were processed), but it allows for a quick search to find a preliminary sample.
	
\begin{figure*}
		\centering
		\includegraphics[width=\textwidth]{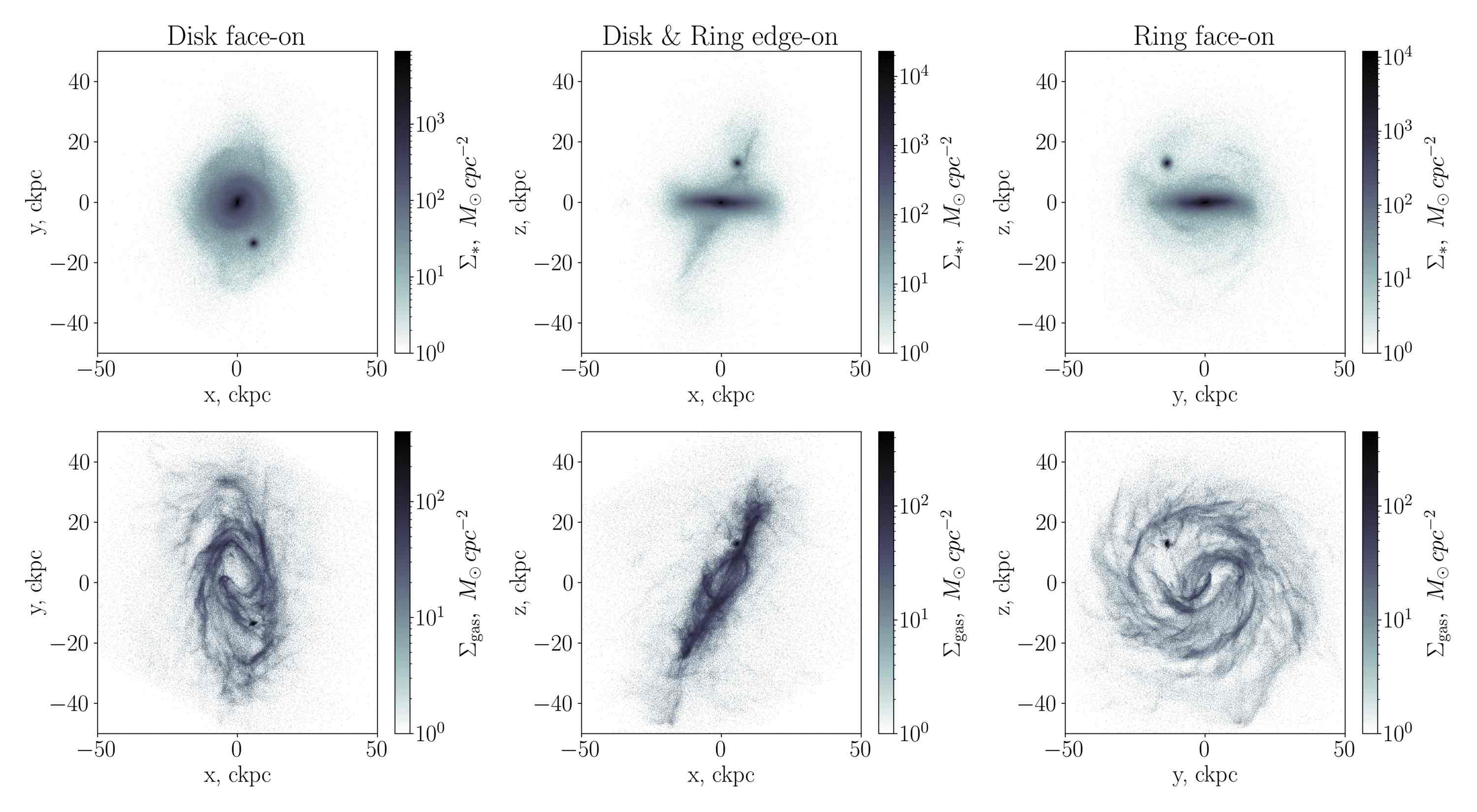}
		\caption{Surface mass density maps of galaxy 501208 (SubhaloID at $z=0$) in three projections. \textit{Top row}: stars, \textit{bottom row}: gas.}
		\label{fig:SMD_map}
	\end{figure*}

	In the next step using \textsc{Sublink} merger trees for each of the 31 candidates, their $z=0$ descendant was determined as the focus of this study are these ``present-day" PRGs. For a more convenient and reliable identification of polar structures each galaxy was centred at 0 and rotated in a following way: $z$-axis was aligned with the angular momentum of the stellar particles inside 2 stellar half-mass radius ($\hmrad$), $x$-axis -- with the angular momentum of the gas particles\footnote{We have found that gas kinematics traces the rotation axis of a polar structure better.} outside $3.5\hmrad$ and the $y$-axes completes the right-hand system. This transformation ensures that all the galaxies have the same orientation: disc lies in the $xy$-plane while the potential polar structure faces $y$ direction. Such limits for determining the desired orientation of the galaxy in space were chosen because they best highlight the main axes of the galaxy. Surface mass density maps (SMDMs) in three projections were created for all candidates in stellar and gas components. \autoref{fig:SMD_map} shows an example of such maps for the galaxy depicted in the middle panel of \autoref{fig:synt_im}.
	
\begin{figure*}
		\centering
		\includegraphics[width=\textwidth]{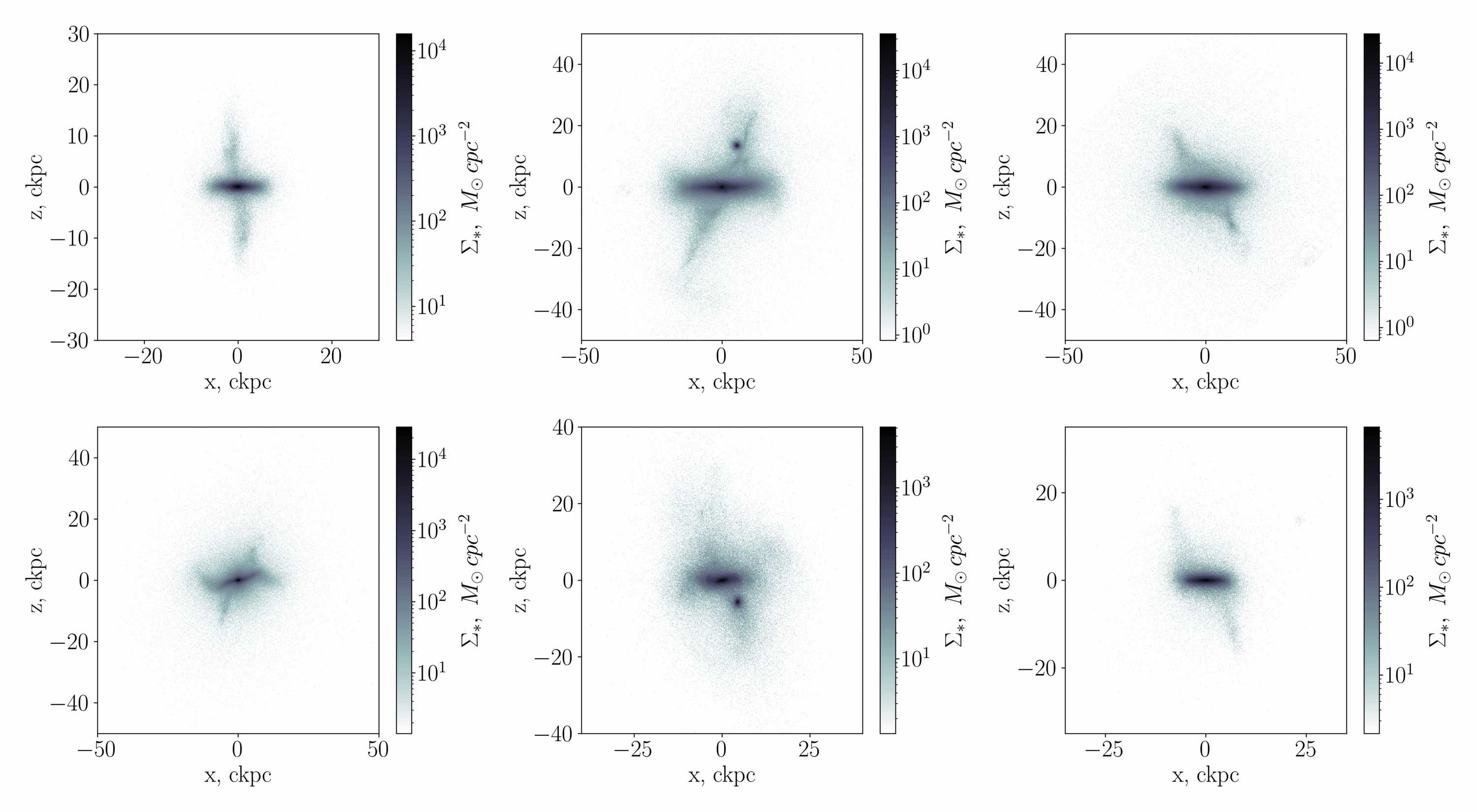}
		\caption{Stellar surface mass density maps of 6 galaxies from the final sample in $xz$-plane. Subhalo ID's are listed in \autoref{tab:main_params} from left to right top to bottom.}
		\label{fig:sample}
	\end{figure*}

	A galaxy was classified as a true PRG if a polar structure was found in the stellar component at $z=0$ (as that is what we see in observations). A total of six galaxies meeting this criterion were found among the sample of 31 candidates. In each case, a gas polar structure was found alongside the stellar ring. Stellar surface mass density maps of six galaxies that make up the final sample are shown in \autoref{fig:sample} in ``disc \& ring edge-on" projection.
	During our search we have also encountered a comparable number of galaxies with gaseous PS but no stellar counterpart. These objects lie outside the scope of this study as we are interested in classical stellar polar rings for which there is a sufficient amount of observational data we could compare our results to. Although a number of exclusively gaseous PS were discovered \citep{Demers2006, Stanonik2009, Deg2023}, this scarce statistics does not allow for any reasonable comparison with simulations. Our finding of as many of these objects as stellar rings imply that there should be a population of hundreds of gaseous polar rings not yet discovered (similar results were obtained by \citealt{Deg2023}).
	
\begin{figure*}
		\centering
		\includegraphics[width=\textwidth]{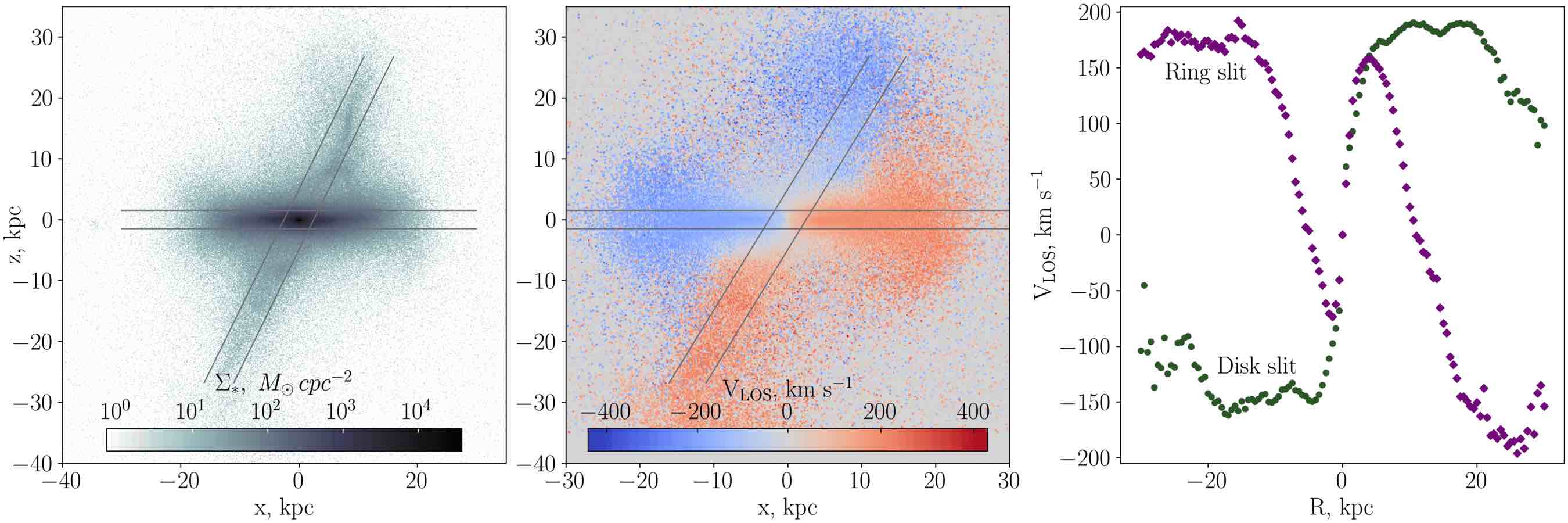}
		\caption{Surface mass density maps (\textit{left panel}), velocity maps (\textit{central panel}) and velocity profiles (\textit{right panel}) for the stellar component of a galaxy 501208. SMD and line-of velocity maps are shown in ``Disk \& Ring edge-on" projection. Line-of-sight velocity profiles are calculated along two slits aligned with the galaxy's disk (green dots) and ring (magenta diamonds) depicted by grey lines on central and left panels.}
		\label{fig:vmap}
	\end{figure*}

	To confirm that found objects consist of two distinct components and justify the correctness of further analysis kinematic maps of six subhalos were inspected. All galaxies indeed demonstrate a typical PRG kinematics with two decoupled rotating subsystems, similar to the observed galaxies with polar rings. An example of a velocity map for one of the galaxies is set out on \autoref{fig:vmap} (similar to Figure 10 in \citealt{Pillepich2019}).
	
	During the search we have found two peculiar galaxies with highly inclined ring structures (Subhalo IDs 117250 and 289385). These galaxies were excluded from consideration as they do not resemble real polar-ring galaxies -- they have significantly higher masses ($M_*\sim 10^{12}\msun$) and reside in a crowded environment compared to PRGs in observations \citep{Savchenko2017}.
	
	\subsection{Kinematic decomposition}\label{subsec:kin_decomp}
	
	In order to better study the properties of polar structures, they need to be separated from the main body of the galaxy. The standard way to distinguish particles from different subsystems is to perform a kinematic decomposition of a galaxy in question. Illustris Supplementary Data Catalogues \citep{Du2019, Du2020} contain very limited results obtained from such decomposition, which also considered only regular subsystems: warm and cold discs, discy bulge, classical bulge and halo. So, a proper kinematic decomposition of selected galaxies is required to separate them into HG and PS.
	
	The classical kinematic decomposition method was presented by \citet{Abadi2003} and allows for an easy determination of disc and spheroid mass fractions. Decomposition is based on a circularity parameter defined as a ratio between the $z$-component of the angular momentum of a particle ($J_z$) and the angular momentum of the particle on a circular orbit having the same binding energy ($J_c$): $\epsilon\equiv J_z/J_c$. The circularity distribution is usually bimodal: disc particles give a sharp peak near $\epsilon\sim1$ while a non-rotating spheroid produces a symmetric bump around $\epsilon=0$. Although this method is useful for the mentioned task, it is not sufficient for a proper classification of each particle.
	
	\citet{Domenech-Moral2012} moved to a 3D phase space adding planar angular momentum ($\boldsymbol{J}_p=\boldsymbol{J}-\boldsymbol{J}_z$) normalised to $J_c$ and binding energy $E$ to circularity. The first new parameter traces out-of-plane motions (which is the main signature of polar structures) and binding energy separates disc components from a more bound bulge. An unsupervised clustering algorithm $k$-means \citep{Scholkopf1998} was employed to distinguish galactic subsystems in this phase space. \citet{Obreja2016} further developed this method by replacing the $k$-means algorithm with an unsupervised machine-learning algorithm, the Gaussian Mixture Model (GMM), and switching to specific angular momentum. The authors discussed the limitations of $k$-means in regards to kinematic decomposition and showed that GMM is a more favourable algorithm for the task. It was later shown \citep{Du2019, Du2020} that a GMM clusterization algorithm in a 3D phase space successfully performs a kinematic decomposition of a large sample of disc galaxies from TNG100 and TNG50 simulations.
	
	This study follows the same approach by performing a GMM clusterization in a 3D phase space. In the first step, the specific angular momenta and binding energy for each particle in a studied galaxy are computed with the help of MORDOR package \citep{MORDOR}. Then three phase space coordinates are calculated: $\jzc,\ \jpc,\ \e$, where $e_\mathrm{max}$ is the energy of the most bound particle. In the second step, we run a GMM clusterization algorithm on stellar particles and gas cells separately using a \textsc{Python} \texttt{scikit-learn} package \citep{scikit-learn}. In order to exclude intruders and kinematic outliers, only particles with $\jzc\in[-1.5, 1.5]$, $\jpc\in[0, 1.5]$ and $\e\in[-1, 0]$ are considered for decomposition, following \citet{Du2019}. As the sole objective of this stage is to separate each galaxy into polar structure and host galaxy, the number of clusters was chosen so as to best distinguish the PS from the rest of the galaxy and often did not match the real number of subsystems.
	
\begin{figure*}
		\centering
		\includegraphics[width=\textwidth]{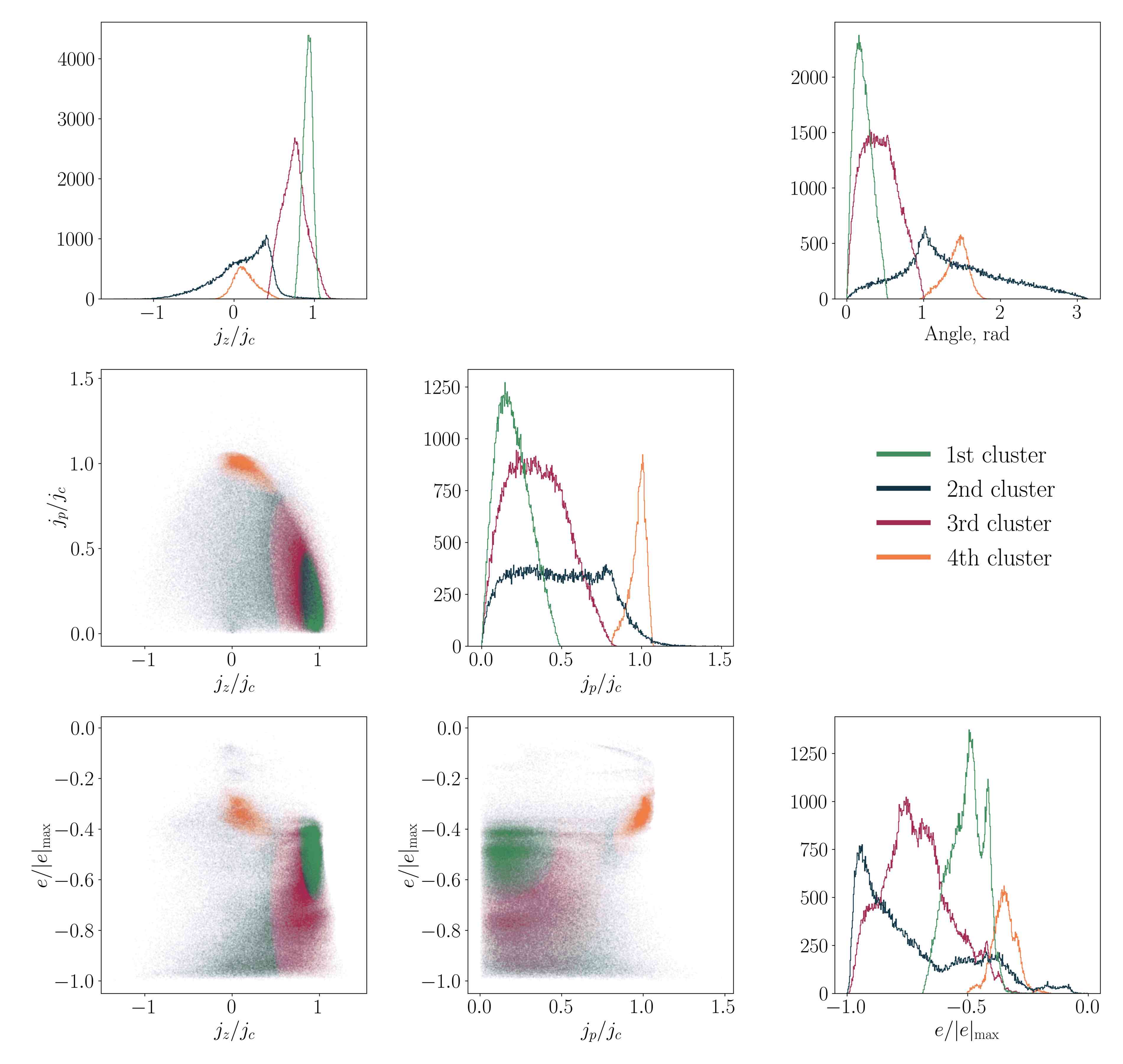}
		\caption{Phase pace of stellar particles from the 595100 galaxy (upper left panel on \autoref{fig:sample}). Panels on the diagonal show the 1D distribution for $\jzc$, $\jpc$ and $\e$. Panels under the diagonal give the 2D distributions of particles belonging to different clusters. The upper right panel depicts the distribution for angle between the $z$-axis and angular momentum of stellar particles from different clusters.}
		\label{fig:phase_space}
	\end{figure*}

\begin{figure*}
		\centering
		\includegraphics[width=\textwidth]{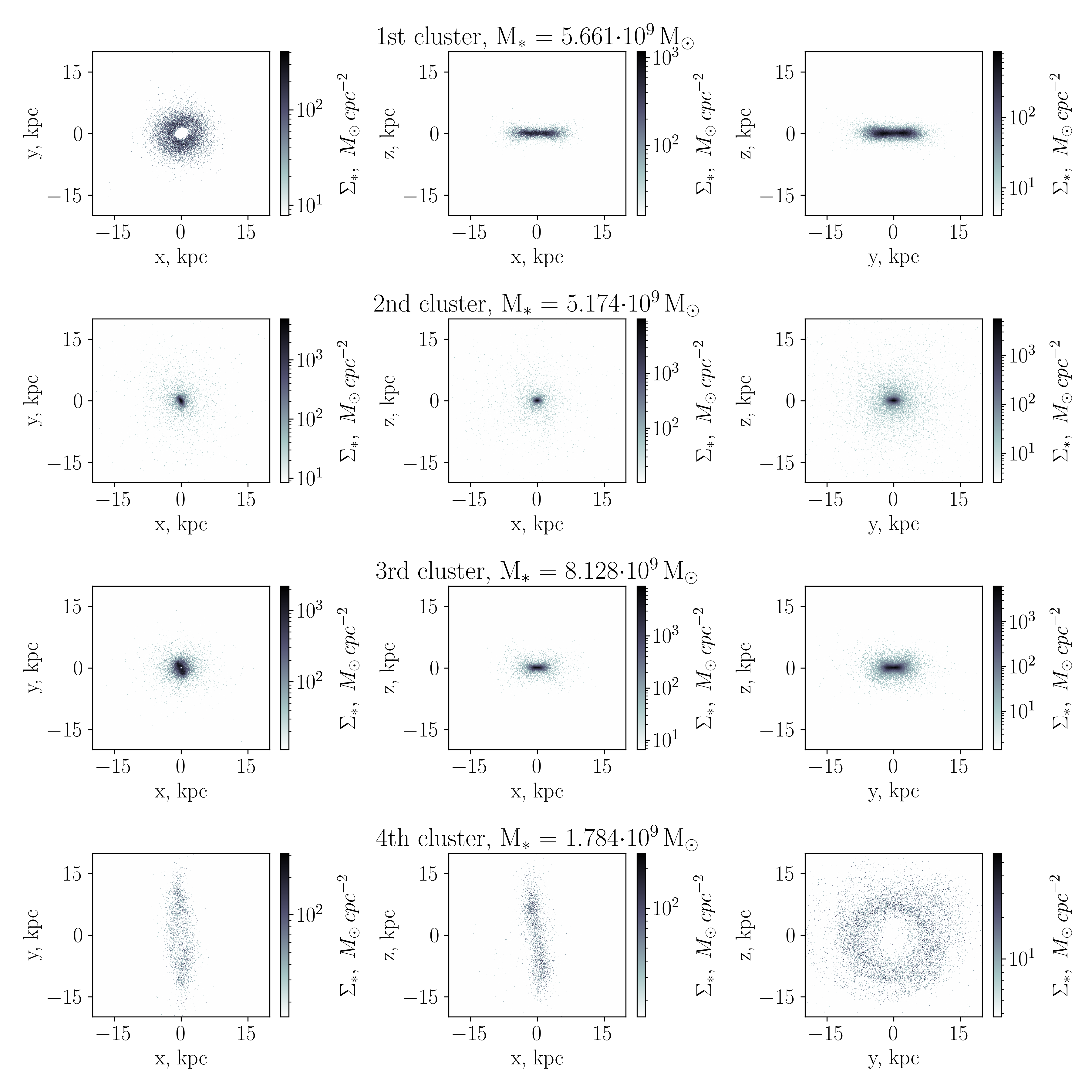}
		\caption{Surface mass density maps of four different subsystems derived during kinematic decomposition of 595100 galaxy. Each row gives maps in three projections (same as in \autoref{fig:SMD_map}) and a total mass of the subsystem. Clusters are easily identified as cold disc, spheroid, warm disc and polar structure (from top to bottom).}
		\label{fig:ups}
	\end{figure*}

	\autoref{fig:phase_space} provides an example of a GMM clusterization in a 3D phase space for stellar particles from one of the galaxies in our final sample. Panels on the diagonal show the distributions of particles by the three kinematic parameters, while panels under the diagonal present three 2D projection of the phase space ($\jpc$ vs. $\jpc$, $\e$ vs. $\jzc$ and $\e$ vs. $\jpc$). In the upper right corner, a distribution for the angle between angular momentum and $z$-axis is set out. The algorithm was run with four clusters depicted by different colours. The distributions allow for an easy interpretation of the physical nature of each cluster. For instance, it is clear that the first cluster with a sharp peak at $\jzc\approx1$ and minimal out-of-plane motions ($\jpc\sim0$) corresponds to the cold disc, while the fourth cluster centred at $\jzc\sim0$ with a narrow maximum at $\jpc\approx1$ is a polar ring which is dominated by out-of-plane motions. Surface mass density maps of each cluster presented in \autoref{fig:ups} confirm this analysis.
	
	The described kinematic decomposition procedure was applied to the gas and stellar components of the 6 galaxies in our sample and all their progenitors up to the moment of the PS formation.
	
	\section{Characteristics of PRGs in TNG50}\label{sec:characteristics}
	
	A plethora of various Supplementary Data Catalogues available for TNG simulations makes it possible to study observable parameters of simulated galaxies and compare them to their real-world counterparts. \citet{Trcka2022} calculated highly realistic multi-wavelength photometry and SEDs for all galaxies with $M_*>10^8\,\msun$ at $z=0$ and $z=0.1$. These values are computed with SKIRT radiative transfer code \citep{SKIRT1, SKIRT2} and SED-fitting tool \textsc{CIGALE} \citep{CIGALE} which account for presence of dust, internal absorption and scattering in galaxies. Spectrophotometric measurements are available for three orientations (face-on, edge-on and random) and for four apertures (10 kpc, 30 kpc, $2\hmrad$, $5\hmrad$). For all six galaxies we determined the orientation which matches the best apparent morphology of observable PRGs (disc\& ring edge-on, typically) by visual inspection of SMDMs in each orientation. We have found that out of four available apertures $r_g\equiv\min\left\{30\,\text{kpc},\ 5r_{1/2,\,*}\right\}$ produces the most realistic cover of the host galaxy and polar structure. To compare with observational data, for each galaxy absolute magnitudes in the $g$ and $r$ SDSS bands were extracted for the determined orientations and apertures.
	
	To compare sizes of the real and simulated PRGs effective radii were chosen, as they are available for both sources. The data for TNG galaxies was extracted from a catalogue of stellar projected sizes originally presented by \citet{Genel2018}. It provides half-light radii for each galaxy with $M_*>3\times10^8\msun$ from TNG50, TNG-100 and TNG-300 in eight photometric bands ($U$, $B$, $V$, $K$, $g$, $r$, $i$, $z$) in five different orientations (one face-on, three edge-on and one random). Two of the edge-on projections correspond to the largest/smallest projected radii for the star-formation distribution within $2\hmrad$ and the third one is chosen randomly. The ``smallest" edge-on orientation was chosen for our purposes.
	
	The main parameters of polar-ring galaxies from TNG50 are set out in \autoref{tab:main_params}: Subhalo ID at $z=0$, absolute magnitude in the $r$ band, the $g-r$ colour, stellar and gas masses within $r_g$, effective radius in the $r$ band and stellar half-mass radius.
	
	\begin{table*}
		\centering
		\caption{Main parameters of polar-ring galaxies from TNG50 simulations.}
		\begin{tabular}{ccccccc}
			\hline
			\multirow{2}{*}{Subhalo ID} & $M_r$, & \multirow{2}{*}{$g-r$} & $\log_{10}M_*$ & $\log_{10}M_\mathrm{gas}$ & $r_e$, & $\hmrad$,\\
			& mag & & $\msun$ & $\msun$ & kpc & kpc \\ \hline
			595100 & -20.58 & 0.56 & 10.29 &  9.62 & 2.24 & 2.22 \\
			501208 & -21.95 & 0.64 & 10.92 & 10.38 & 6.50 & 5.90 \\
			483594 & -21.63 & 0.72 & 10.93 &  9.85 & 3.69 & 3.54 \\
			514272 & -21.01 & 0.61 & 10.54 &  9.95 & 4.31 & 3.95 \\
			571908 & -21.04 & 0.52 & 10.38 & 10.31 & 4.39 & 6.01 \\
			428178 & -20.33 & 0.58 & 10.27 &  9.86 & 2.59 & 3.24 \\ \hline
		\end{tabular}
		\label{tab:main_params}
	\end{table*}
	
\begin{figure}
		\centering
		\includegraphics[width=0.5\textwidth]{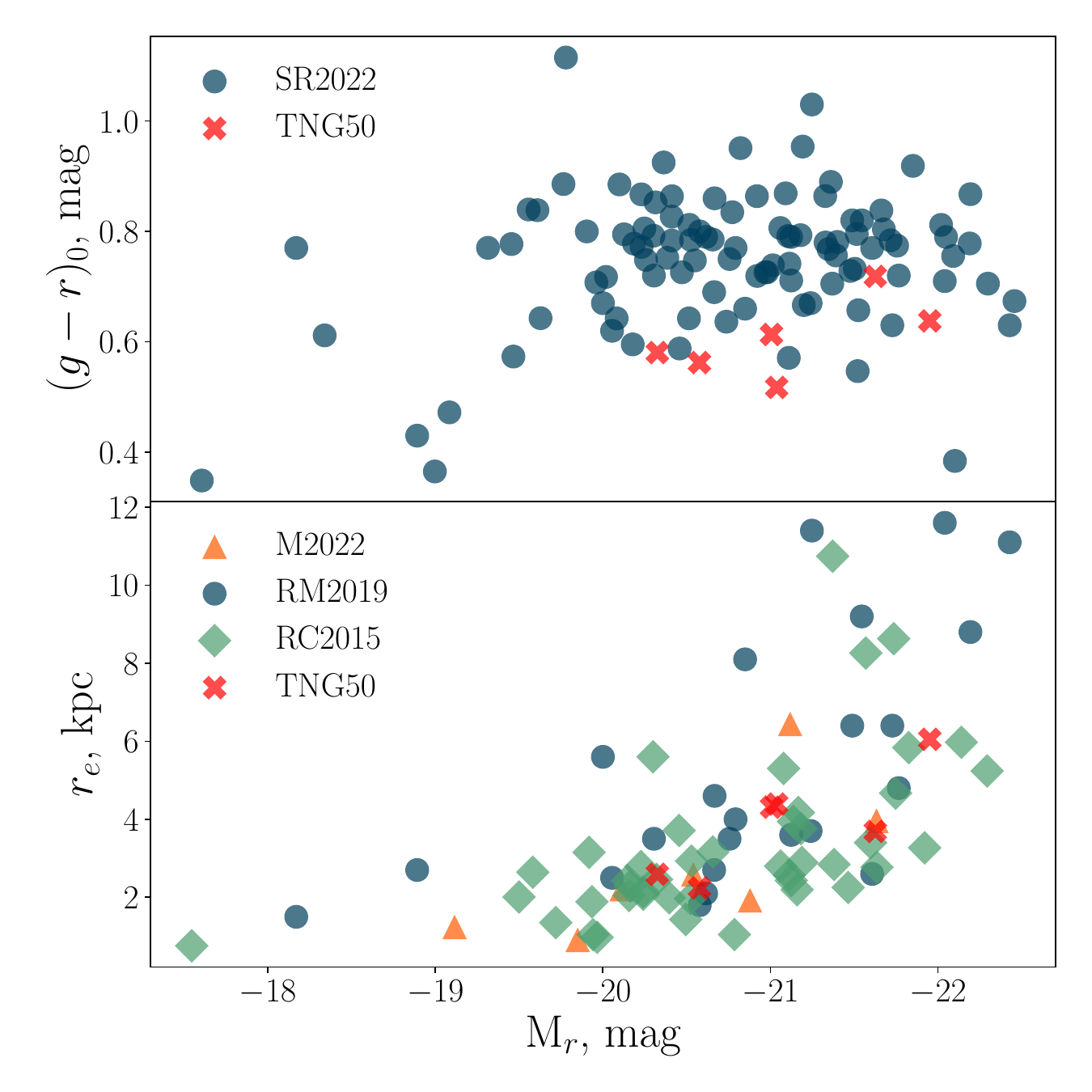}
		\caption{Comparison of the fundamental parameters of the real (blue circles, green diamonds and orange triangles) and simulated (red crosses) PRGs. \textit{Top:} colour-magnitude diagram for the observational sample from \citet{Smirnov2022} (SR2022, blue circles) and our TNG50 sample (red crosses). \textit{Bottom:} distribution of the PRGs from observations and TNG50 in the $M_r-r_e$ plane: observations -- \citet{rm2019} (RM2019, blue circles), centrals from \citet{Reshetnikov2015} (RC2015, green diamonds) and \citet{Mosenkov2022} (M2022, orange triangles), TNG50 -- this paper (red crosses).}
		\label{fig:comparison}
	\end{figure}

	\autoref{fig:comparison} compares the main characteristics of the polar-ring galaxies from observations and TNG50. The upper panel shows a distribution of galaxies on a colour-magnitude diagram. The observational sample is taken from \citet{Smirnov2022} (SR2022) and contains either kinematically confirmed polar-ring galaxies or objects with a very prominent PRG morphology. As follows from the chart, the simulated galaxies have similar luminosities to the real PRGs but bluer colours on average. The absence of low luminosity objects can possibly be attributed to the selection effect: low-luminosity (and, consequently low-mass) galaxies are probably underrepresented in the synthetic image sample compared to more massive galaxies. The colour discrepancy could be, at least partially, explained by geometric effects. In most cases none of three available photometry orientations provides a ``disk\&ring edge-on" view of the galaxy, which is the most common projection in observational data due to selection effects (see \citealt{whit1990}). This inconsistency means that the light emitted by the ring in simulations experiences less internal extinction compared to real observations. Difference of mean colours of observed and simulated PRGs might be considered as a rough estimate of internal reddening of the polar rings. In our case it amounts to 0.14 magnitudes which is consistent with observations of galaxies at different inclinations \citep{Masters2010}.
	
	The distribution of the polar-ring galaxies from TNG50 and observations in the $M_r-r_e$ plane is presented in the bottom panel of \autoref{fig:comparison}. Observational data was taken from \citet{rm2019} (blue circles), \citet{Mosenkov2022} (orange triangles) and \citet{Reshetnikov2015} (green diamonds), where the last two sources contain only effective radius measurements for the host galaxies. As can be seen from the figure, simulated galaxies reside in the same area as the observed PRGs and are consistent with them in the size distribution.
	
	\subsection{Subsystem parameters}
	
	Having performed kinematic decomposition of all galaxies from our sample, we are able to access main properties of polar structures and host galaxies separately. The estimate of integral luminosity of each subsystem was calculated by summing the luminosities of all stellar particles within $r_g$. This method provides only an upper limit on the total luminosity and a lower limit on the galaxy's colour as it does not account for absorption or obscuration in the system. Howether, such an approach provides a reasonable estimate which can be compared to the observational data. The characteristics of polar structures and host galaxies are presented in \autoref{tab:decomp_pars}, it lists absolute magnitude in the $r$ band, $g-r$ colour, and a stellar mass within $r_g$ for each subsystem as well as a ring-to-host luminosity and stellar mass ratio. Within the limitations of our method calculated luminosities and colours of the simulated PRGs are in agreement with the with observational data: host galaxies have red colours and sub-$M^*$ luminosities ($<M^\mathrm{HG}_r>=-20.68\,$mag, $<(g-r)^\mathrm{HG}>=0.74$) in contrast to faint and blue polar structures ($<M^\mathrm{PS}_r>=-18.94\,$mag, $<(g-r)^\mathrm{HG}>=0.61$) \citep{Reshetnikov2015, Mosenkov2022}.
	
	\begin{table*}
		\centering
		\caption{General characteristics of PRGs' subsystems in TNG50.}
		\begin{tabular}{ccccccccc}
			\hline
			& \multicolumn{3}{c}{Polar Structure} & \multicolumn{3}{c}{Host galaxy} &  &  \\
			Subhalo ID & $M_r$ & $g-r$ & $\log_{10}M^\text{PS}_*$ & $M_r$ & $g-r$ & $\log_{10}M^\text{HG}_*$ & $L^\text{PS}/L^\text{HG}$ & $M_*^\text{PS}/M_*^\text{HG}$\\ \hline
			595100 & -19.69 & 0.12 & 9.25  & -20.44 & 0.53 & 10.28 & 0.50 & 0.10 \\
			501208 & -21.07 & 0.22 & 10.09 & -21.52 & 0.78 & 10.87 & 0.66 & 0.17 \\
			483594 & -19.84 & 0.14 & 9.39  & -21.52 & 0.78 & 10.95 & 0.21 & 0.03 \\
			514272 & -19.84 & 0.08 & 9.29  & -20.93 & 0.70 & 10.56 & 0.37 & 0.05 \\
			571908 & -20.05 & 0.22 & 9.72  & -20.97 & 0.48 & 10.30 & 0.43 & 0.27 \\
			428178 & -18.96 & 0.10 & 8.99  & -20.62 & 0.53 & 10.26 & 0.22 & 0.05 \\ \hline
		\end{tabular}
		\label{tab:decomp_pars}
	\end{table*}
	
\begin{figure}
		\centering
		\includegraphics[width=0.5\textwidth]{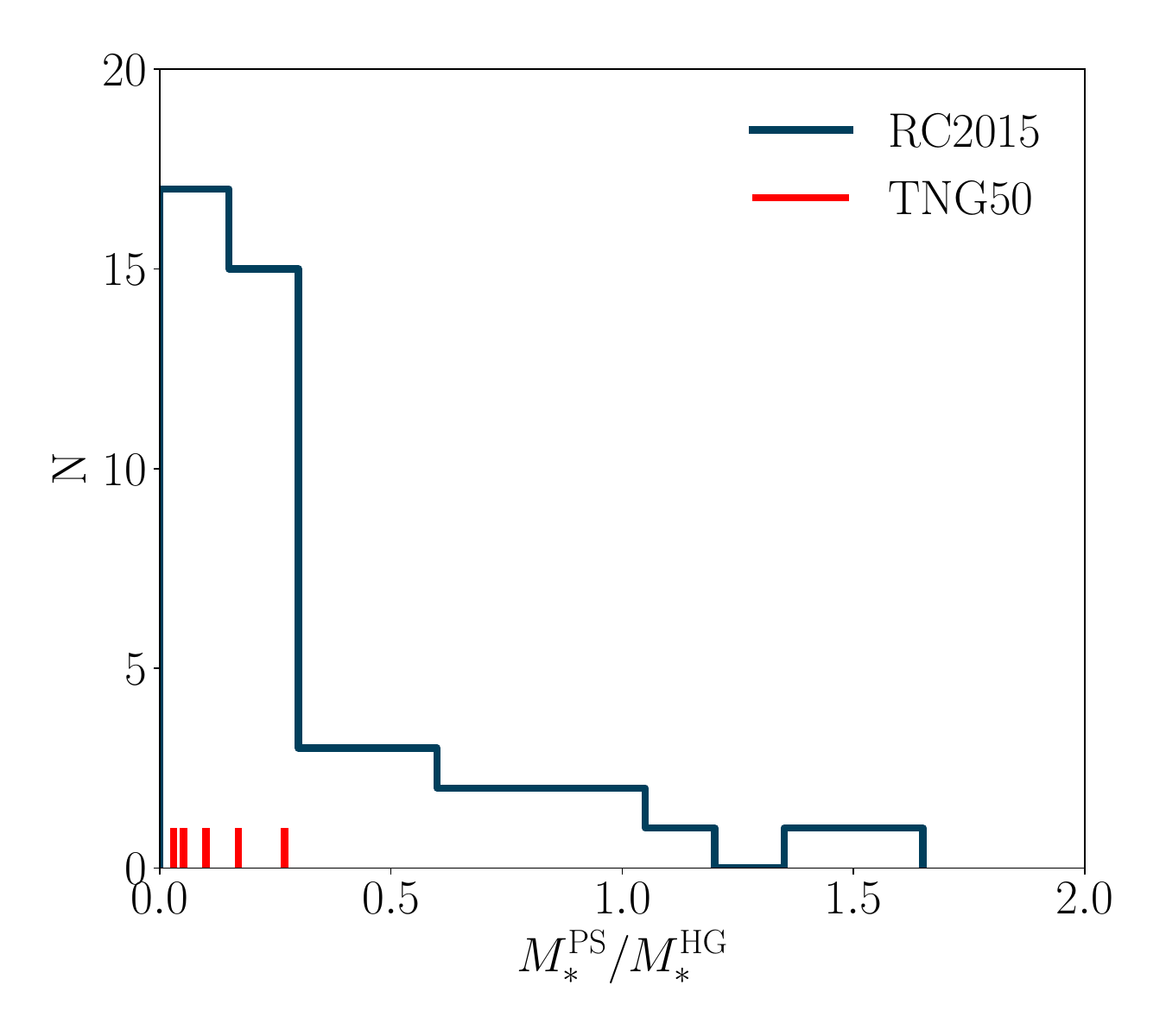}
		\caption{Distribution of observed and simulated PRGs over the ratio between PS and HG stellar masses. TNG50 sample is shown in red with exact values given in \autoref{tab:decomp_pars}, observational data is taken from \citet{Reshetnikov2015} (RC2015).}
		\label{fig:ratio_dist}
	\end{figure}

	As can be seen from \autoref{fig:ratio_dist} and last column of \autoref{tab:decomp_pars}, Illustris PRGs have relatively low-mass rings compared to their host galaxies. The absence of massive rings may be due to the scarce statistics of objects in a limited volume of TNG50 simulations or the formation scenario of the sample galaxies (see \autoref{sec:formation}). Nonetheless, these results match the observational trend that most PRGs are dominated by the host galaxies ($<M_*^\text{r}/M_*^\text{h}>=0.3$).
	
	We have also computed radii of polar structures around the sample galaxies, defined as
	
	\begin{equation}
		R=\int_0^{10\,\hmrad}r^2\rho(r)\,\dd r/\int_0^{10\,\hmrad}r\rho(r)\,\dd r,
	\end{equation}
	where $\rho(r)$ is the density profile of the ring. The average radius of the rings in our sample is $14.8\pm 4.3$ kpc, which, taking into account the differences in the method, is in good agreement with observations \citep{rm2019}.
	
	\section{Formation of PRGs}\label{sec:formation}
	
\begin{figure*}
		\centering
		\includegraphics[width=0.5\textwidth]{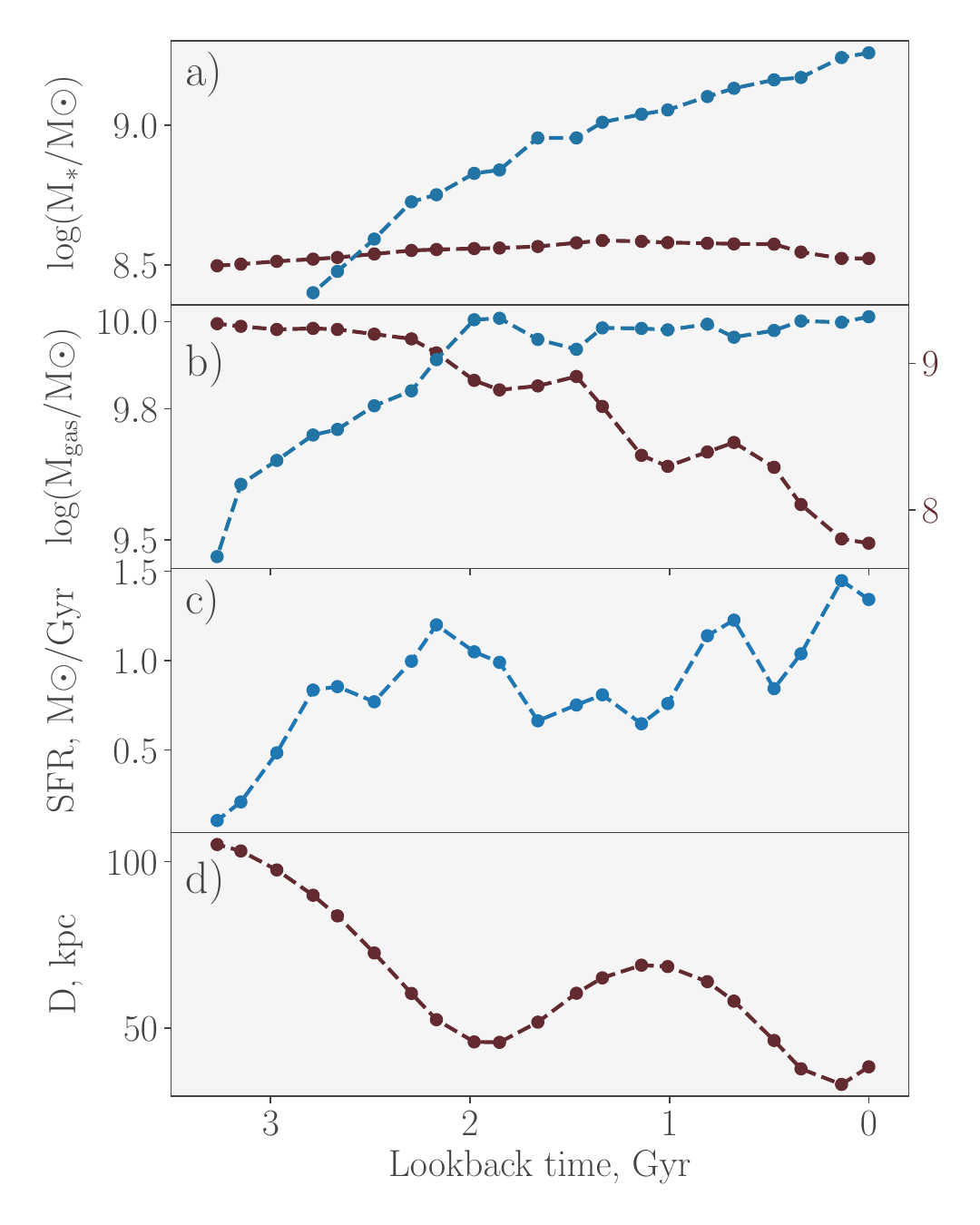}%
		\includegraphics[width=0.5\textwidth]{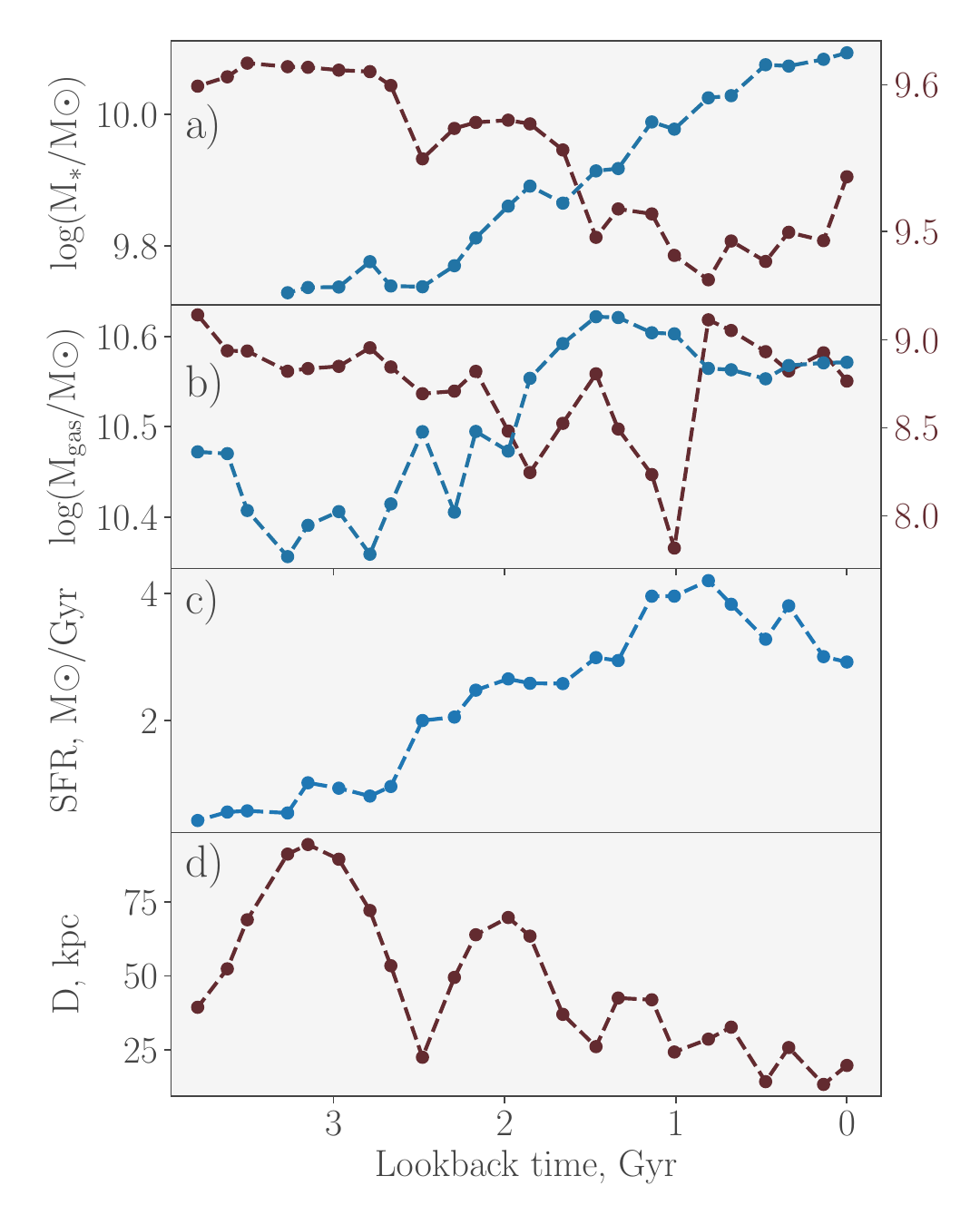}
		\caption{Evolution of main parameters of polar structure (blue) and galaxy companion (red) of Subhalo 595100 (\textit{left}) and 501208 (\textit{right}): (a) stellar mass, (b) gas mass, (c) PS star-formation rate and (d) distance between satellite and host galaxy. Ticks on the right show the scale for the properties of satellites in cases where there is a big difference between values for the PS and satellites. Analogous figures for other Subhalos are presented in \autoref{fig:history_appendix}.}
		\label{fig:history}
	\end{figure*}

	In the previous section, it has been shown that not only morphology but also characteristics of the found subhalos are very similar to those of real polar-ring galaxies, so their history can shed light on the formation of the PRGs. Using \textsc{Sublink} merger trees available for all TNG50 subhalos, we have traced main branch progenitors of every galaxy from the sample up to the moment of the polar structure formation in the stellar and gas components. At each snapshot, the progenitors were decomposed the same way as described in Section~\ref{subsec:kin_decomp} to study the evolution of polar structures with time. Close inspection of the SMDMs revealed that in each case formation of each polar structure was associated with a close interaction between the host galaxy and its companion or satellite. \autoref{fig:history} shows the evolution of the main parameters (stellar mass, gas mass, PS star-formation rate and satellite's galactocentric distance) of the galaxy companion and the polar structure during its formation for two galaxies from our sample.
	
	\autoref{tab:sat} provides data on the stellar and gas masses of the galaxy companions associated with formation of polar structures. For each TNG50 polar-ring galaxy it gives Subhalo IDs of relevant satellites, their stellar and gas masses, as well as the mass ratios between the host and the satellite: $q_*$ for stellar masses, $q_\textrm{gas}$ -- for gas. As host galaxies primarily dominate their halos, most of the intergalactic medium and sometimes satellite's gas periphery is attributed to them which makes it hard to compare their gas masses with the satellites'. To circumvent this obstacle, the  masses are calculated within $r_g$ for each galaxy (satellite or host). The values are presented for snapshots before the interaction occurs in each component. Note that the interaction in different components often starts at different times (see below), so the stellar and gas masses may not correspond to the same snapshot. As can be seen from the table, the relative gas content of the companion galaxies associated with the formation of the polar structures is, on average, higher than that of the PRGs as a whole.
	
	\begin{table}
		\centering
		\caption{Masses of the companion galaxies of PRGs from TNG50.}
		\begin{tabular}{cccccc}
			\hline
			Subhalo ID & Satellite ID & $\log_{10}M_*$ & $\log_{10}M_\text{gas}$ & $q_*$ & $q_\text{gas}$ \\ \hline
			\multirow[t]{2}{*}{483594} & 483600 &  9.11 &  8.89 &  69 &  15 \\ 
			& 483599 &  8.49 &  7.99 & 288 & 123 \\ 
			\multirow[t]{2}{*}{571908} & 571910 &  9.48 &  9.64 &   8 &   7 \\ 
			& 571909 &  9.08 &  9.01 &  20 &  31 \\ 
			595100                     & 595101 &  8.46 &  9.51 &  71 &   3 \\ 
			514272                     & 514273 &  10.2 &  10.1 &   2 &   1 \\ 
			501208                     & 501209 &  9.54 &  9.33 &  20 &   7 \\
			428178                     & 428181 &  9.20 &  9.95 &  12 &   1 \\ \hline
		\end{tabular}
		\label{tab:sat}
	\end{table}
	
\begin{figure*}
		\centering
		\includegraphics[width=\textwidth]{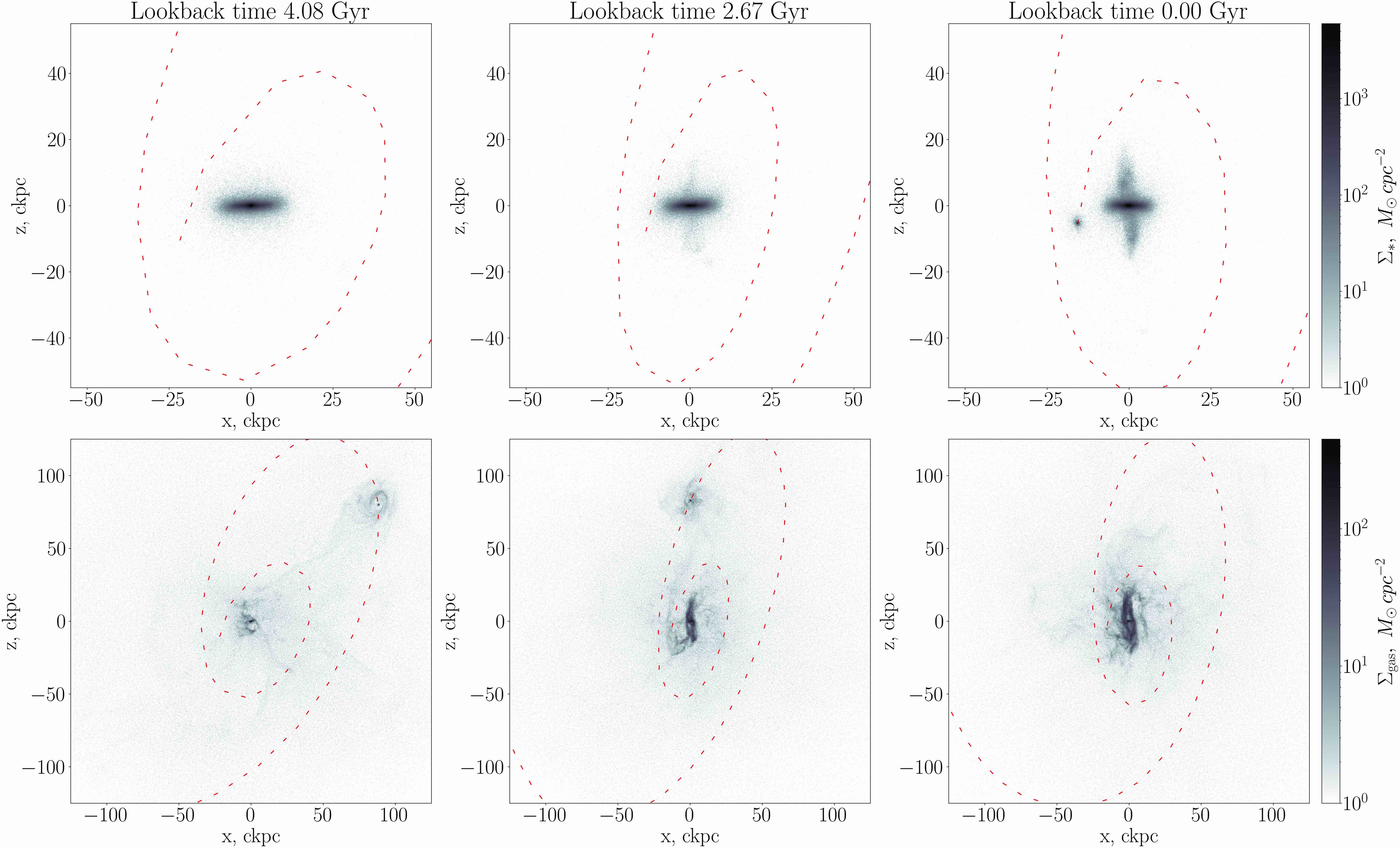}
		\caption{Surface mass density maps of a polar structure formation around Subhalo 595100 due to gas accretion from a satellite. Each row depicts a different galaxy's component (\textit{top} -- stellar, \textit{bottom} -- gas) and columns a different snapshot. The left column shows the early stages of a gas accretion while the middle one presents the beginning of the stellar ring growth due to star formation. The dashed red lines represent the satellite's orbit.}
		\label{fig:formation_acc}
	\end{figure*}

\begin{figure*}
		\centering
		\includegraphics[width=\textwidth]{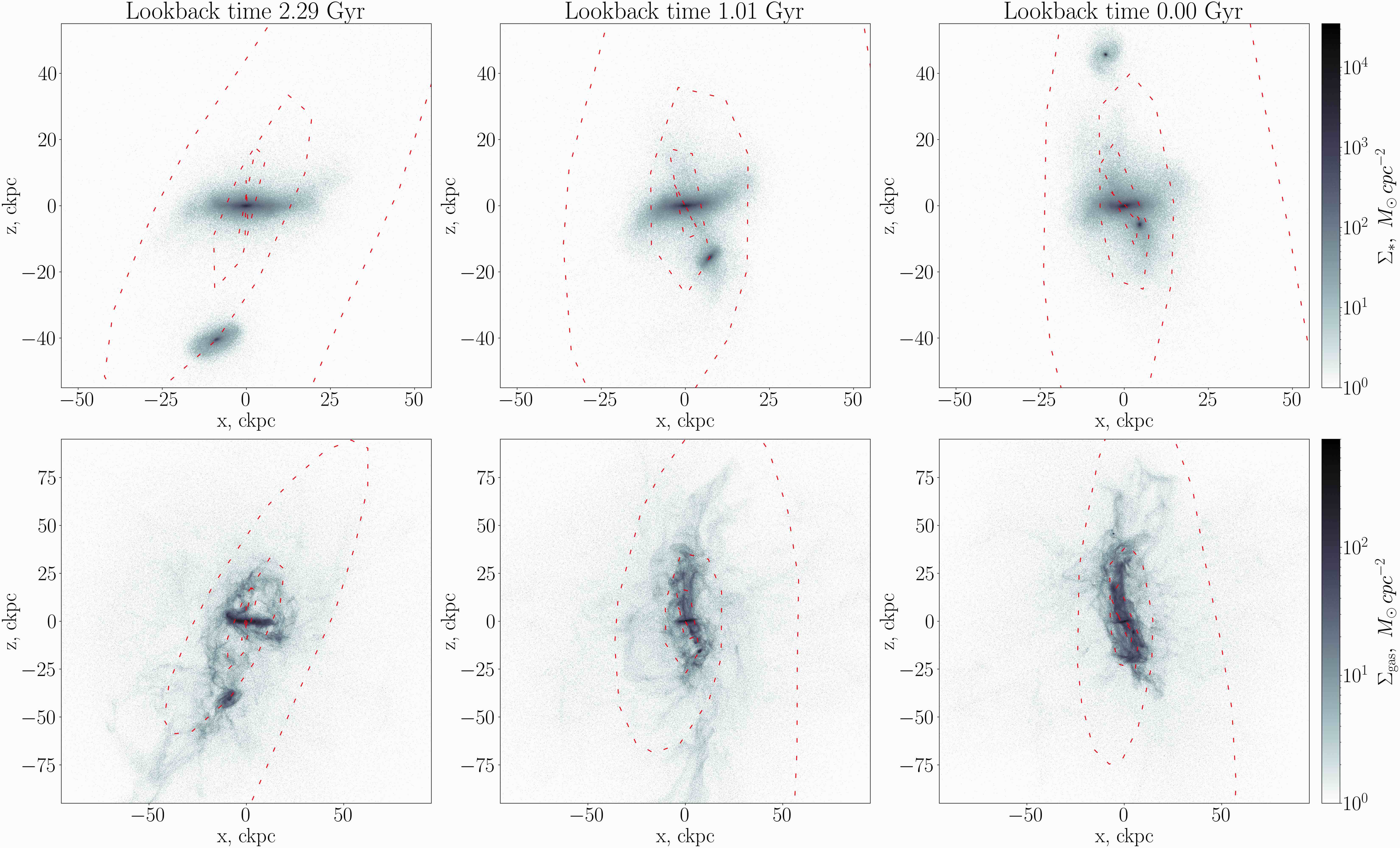}
		\caption{Surface mass density maps of a polar structure formation around Subhalo 571908 during satellite disruption. Each row depicts a different galaxy's component (\textit{top} -- stellar, \textit{bottom} -- gas) and columns a different snapshot. The left column shows the formation of a gas ring while the middle one presents the early stages of interaction in the stellar component. The dashed red lines represent the satellite's orbit.}
		\label{fig:formation_dis}
	\end{figure*}

	Below we provide a short description of the formation process for each galaxy in our sample. Two examples of polar ring formation are presented in \autoref{fig:formation_acc} and \autoref{fig:formation_dis}.
	
	\begin{description}
		\item Subhalo 595100 shown in \autoref{fig:formation_acc}: an isolated galaxy with a satellite. Moving from the apocenter at $z=0.46$, the satellite reaches a sufficiently close distance to start accretion from its periphery onto the central galaxy. Satellite's orbit is oriented almost perpendicular to the central galaxy, so the gas funnels into the polar plane of the host galaxy. At $z=0.26$, the accreted gas forms a polar disc where stars start to form. A stellar ring slowly grows due to this star-formation from this point on. The gas disc of the central galaxy survives the accretion up to $z=0$. At $z=0.1$, moving towards the pericenter, the satellite approaches close enough to start losing some of its stellar component which settles into the polar ring.
		
		\item Subhalo 514272: an isolated galaxy with two companions. At $z=0.26$ during a passage of the pericenter by a more massive companion gas accretion  starts onto the early-type central galaxy. Its orbit lies in a polar plane, so during the passage a transient gas ring occurs with $R\lesssim10$ kpc. During the second passage of the pericenter accretion rates seemingly increase and a proper gas ring is formed with $R\approx15$ kpc. Due to the relatively high pericentric distance ($\sim115$ kpc), there is no accretion of the stellar component from the companion, but the host's disc warps significantly. Both passages are associated with an increase of the star formation rate in accreted gas which ultimately leads to the formation of a stellar ring. The ring's survival in the future remains unclear -- tidal forces may destroy it during the next pericenter passage.
		
		\item Subhalo 428178: galaxy with a satellite in a pair ($D=1.3$ Mpc, $q^*=1/1.5$). Tidal gas accretion from the satellite's outskirts and the IGM starts at $z=0.3$ when it approaches the host galaxy moving from the apocenter. Due to its angular momentum gas settles into a plane inclined at an angle of $\sim75^\circ$ to the host galaxy. There, accreted gas forms a spiralled disc which attaches to a gas disc of the central galaxy. The stellar ring gradually grows due to the star formation induced in the accreted gas. By $z=0$, the satellite is close enough to start stellar accretion.
		
		\item Subhalo 571908 shown in \autoref{fig:formation_dis}: galaxy with two satellites. A closer and more massive satellite (571910) steadily approaches the central galaxy in the near-polar plane. At $z\approx0.27$, tidal forces from the host are sufficient to begin gas accretion from the satellite's periphery. At $z=0.17$ tidal disruption of its stellar component begins as the distance has shortened up to 40 kpc. By this time, the satellite had lost most of its gas, which formed a massive and extended disc in the polar plane of the host galaxy. The gas disc of the central galaxy gradually shrinks but survives by $z=0$. Starting with $z\approx0.1$ interaction with the second satellite (571909) begins, which rapidly loses gas and, when approaching its pericenter at $z=0.02$, a decrease in the stellar mass is also observed.
		
		\item Subhalo 501208: galaxy with a system of satellites. Moving along a highly inclined orbit at $z=0.36$, one of the satellites comes close enough to start gas accretion from its periphery onto the massive central galaxy. The stripped gas settles into a disc in a near-polar plane and starts forming stars. During the next pericenter passage at $z=0.2$ ($D=20$ kpc), the satellite drags with it the remainder of the host's gas disc which further fuels the formation of the polar structure. At the same time, the satellite begins to experience tidal disruption of its stellar component feeding the forming stellar ring. It is clear that in the future the satellite will be fully absorbed by the polar structure.
		
		Subhalo 483594: galaxy with a system of satellites. Starting at $z=1$ when a more massive satellite passes the pericenter, gas accretion onto the central galaxy starts. Due to the orbital momentum of the satellite the accreted material falls into the polar plane where by $z = 0.44$ a gas disc is formed and star formation starts. At the same, time with each passage of the pericenter the satellite loses a fraction of its stellar mass, which also settles in the forming polar ring. A less massive satellite moves along a more eccentric orbit, and for a long time is gradually losing its gas to the growing polar structure. Approaching $D=50$ kpc at $z\approx0$, it quickly loses a significant part of the remaining gas and a small fraction of stars, triggering a new burst of star formation in the ring.
	\end{description}
	
	Based on the above analysis, we have found that the sample galaxies could be divided into two (possibly evolutionarily related) groups based on their formation scenario. The ring may be formed due to star formation in the gaseous polar structure accreted from the satellite. Or the ring is formed mainly due to the tidal disruption of the stellar component of the satellite. Both formation routes are subcases of the accretion scenario. It is noteworthy that we have not found a single galaxy formed during major mergers or cold filament accretion.
	
	\citet{bc2003} showed that the probability of two merging spirals forming a polar-ring galaxy is about 1\%. Knowing the number of large mergers ($q_*\gtrsim0.7$) in simulations, it is possible to estimate the expected number of PRGs formed during the merging scenario. Thus, among a sample of large Milky Way/M31 type galaxies from TNG50 \citep{SotilloRamos2022}, the expected number of polar structures with collisional origin is $\lesssim0.5$; extrapolation of the frequency of large mergers in TNG100 (\citealt{Bickley2021, ByrneMamahit2023}) to a smaller volume of TNG50, also gives the expected number of ``collisional" PRGs less than 1. So the absence of polar rings formed during galaxy mergers in our sample is easily explained in the existing paradigm by the insufficient volume of the simulated area.
	
	The ages of the stellar polar structures are presented in \autoref{tab:age}: $t_\text{form}$ is the lookback time of a snapshot where the stellar PS was firstly extracted from the galaxy by kinematic decomposition and $<t>_m$ is the mass-weighted mean age of the ring's stellar population. The data is consistent with observations which indicate that polar rings are relatively young subsystems \citep{Reshetnikov1994, Iodice2002, Iodice2002b}. As can be expected, the stellar populations of PS formed during satellite disruption are, on average, older than their star formation-driven counterparts.
	
	\begin{table}
		\centering
		\caption{Ages of the stellar polar rings in our sample.}
		\begin{tabular}{ccc}
			\hline
			Subhalo ID & $t_\text{form}$, Gyr & $<t>_m,$ Gyr \\ \hline
			595100 & 2.79 & 1.24 \\
			514272 & 1.01 & 0.69 \\
			428178 & 1.66 & 1.17 \\
			571908 & 2.17 & 4.65 \\
			501208 & 3.27 & 2.28 \\
			483594 & 4.74 & 1.95 \\ \hline
		\end{tabular}
		\label{tab:age}
	\end{table}
	
	The relatively small ages of the polar structures in TNG50 indicate that the spatial environment (interactions, close encounters with other galaxies, disruption of satellites), apparently, limits the lifetime of such structures. (It is noteworthy that we did not find any PRGs in which the ring was formed at $z>0.5$.) This is consistent with the findings of \citet{Savchenko2017}, where the authors found that on scales of $\sim100$ kpc the spatial environment of the PRGs is, on average, less crowded compared to the ``normal" galaxies.
	
	\subsection{Ring inclination}
	
	Although rings are relatively young structures, they exist on timescales of billions of years where the effects of internal dynamics play an important role. One of the most important markers of ring's stability is the evolution of its inclination. 
	
\begin{figure}
		\centering
		\includegraphics[width=0.5\textwidth]{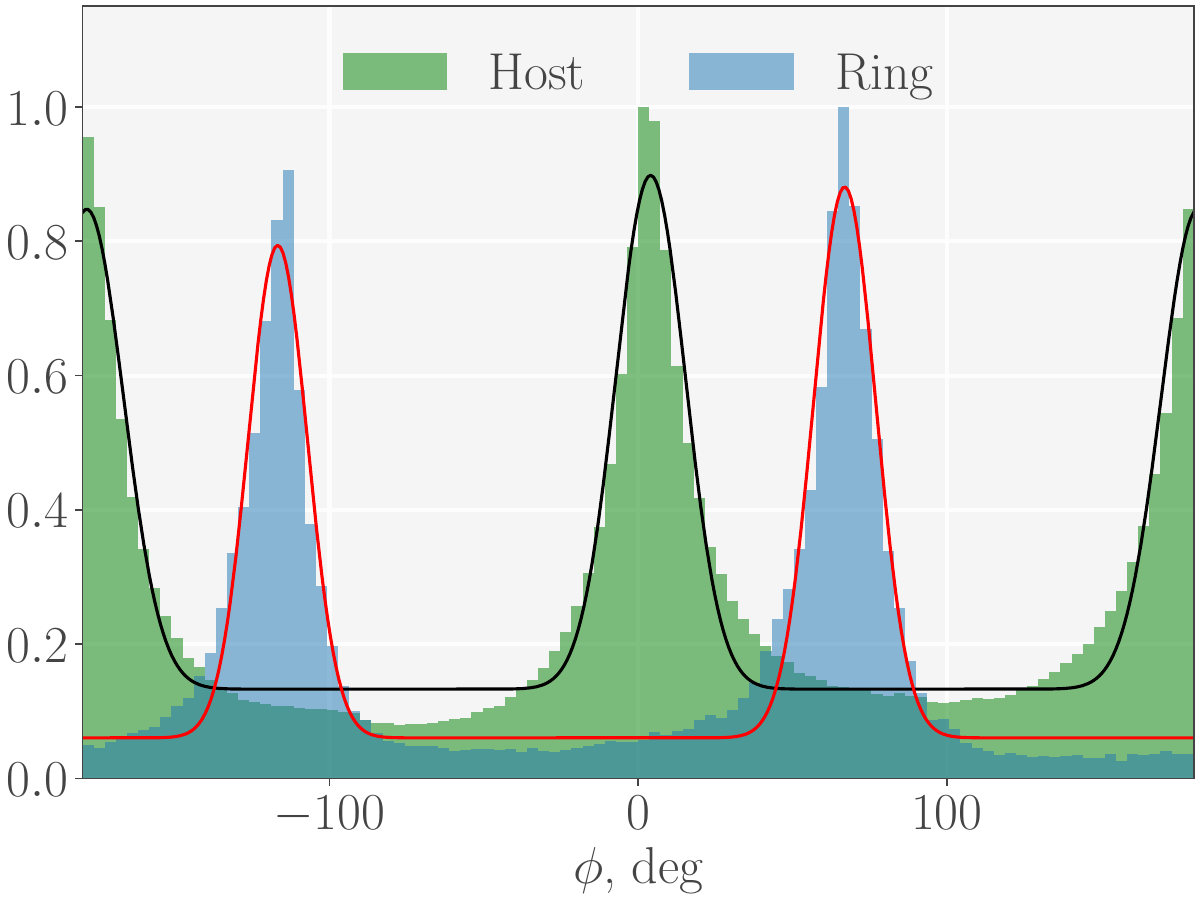}
		\caption{The distribution of stellar particles' polar angle for the ring (blue) and host galaxy (green) of Subhalo 501208. Solid curves show the fitted combination of gaussians for both components.}
		\label{fig:inc_det}
	\end{figure}

	To determine the ring's inclination, we have plotted the distribution of polar angles of the stellar particles in the $xz$-plane (disc \& ring edge-on). The distribution has two peaks corresponding to two sides of the ring with a uniform spread between them. A combination of two scaled up gaussians was fitted to the observed distribution to locate the exact position of both peaks. The inclination is calculated as an error-weighted mean of the centres of the peaks minus 90 degrees, which gives the angle $i$ between the polar and the ring planes. To check that this measurement is not contaminated by the inaccuracy of the galaxy plane position, the same procedure was carried out for the host galaxy as well. In most cases the host's disc lies within $2^\circ$ of the central plane. \autoref{fig:inc_det} shows the distribution of polar angles $\phi$ of 501208 ring's and host's stellar particles and the fitted gaussians for each subsystem.
	
\begin{figure}
		\centering
		\includegraphics[width=0.5\textwidth]{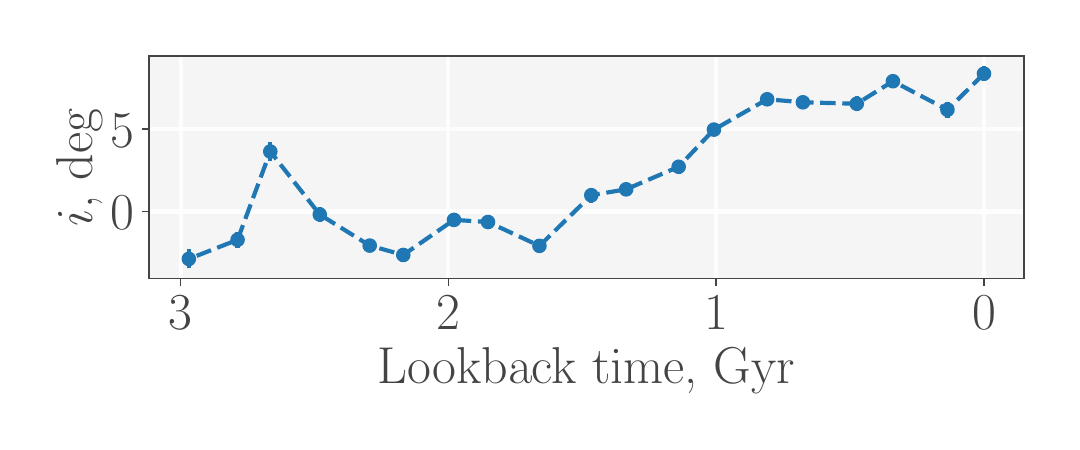}\\
		\includegraphics[width=0.5\textwidth]{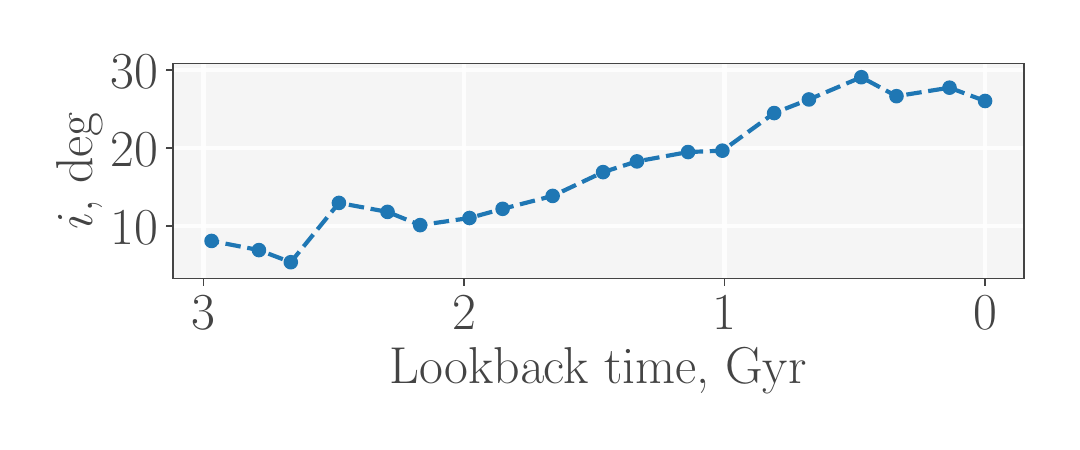}
		\caption{Evolution of the angle between the stellar ring and the polar plane with cosmic time for two galaxies from our sample: \textit{top} -- 595100, \textit{bottom} -- 501208. Analogous figures for other Subhalos are presented in \autoref{fig:inc_appendix}.}
		\label{fig:inc}
	\end{figure}

	This procedure was carried out for all ring hosting progenitors of each galaxy in our sample. The two examples are set out on \autoref{fig:inc}. The two most massive galaxies (Subhalo 501208 and 483594) show a smooth and steady decrease of the rings' inclination from an almost orthogonal position with a rate of $\approx8^\circ/$Gyr. 
	
\begin{figure}
		\centering
		\includegraphics[width=0.5\textwidth]{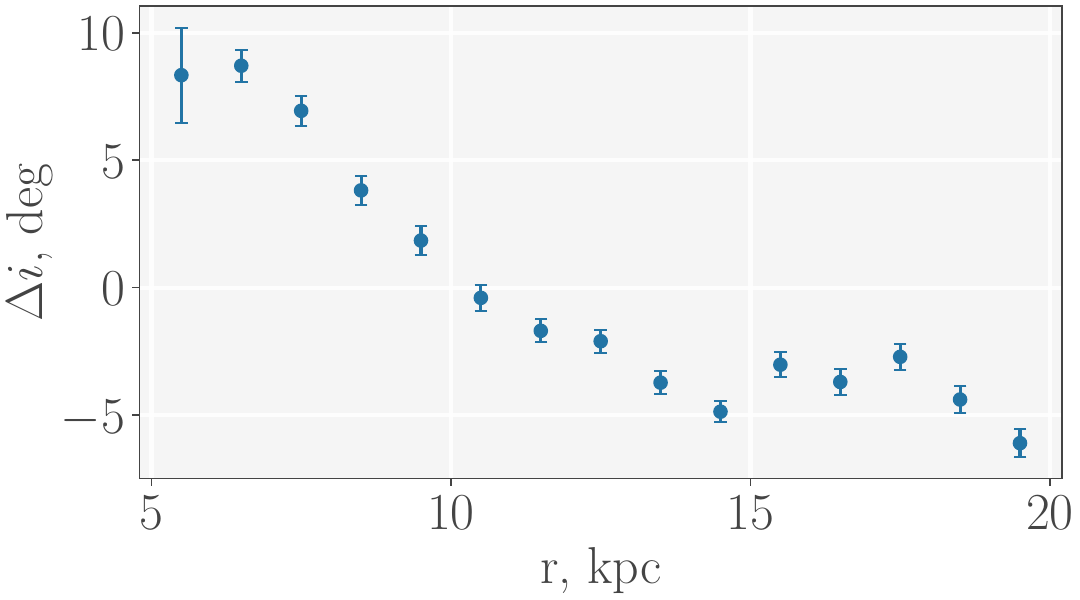}
		\caption{Shape profile of polar ring around galaxy 595100 with a systematic inclination of 3.4$^\circ$. Signature of an upward-warping is clearly seen.}
		\label{fig:warp}
	\end{figure}

	With a similar approach we can measure the ring's shape. To do this, we compute the inclination of the ring's stellar particles in concentric annuli in the same way as described above. Averaging both halves of the ring and subtracting the systematic inclination of the ring as a whole gives us what we call a ``shape profile" of the ring. The shape profile of a straight ring is just a horizontal line centred at zero, while for a warped one it would show systematic differences of opposite signs in the inner and outer parts. Such analysis was performed for all $z=0$ galaxies except for subhalo 571908 where the PS is just forming and has not settled into a relaxed state yet. We report that all studied rings show warps signatures in their shape profiles, which is expected for self-gravitating polar rings. The most vivid example of our findings -- PS around subhalo 595100 -- is set out in \autoref{fig:warp}. The ring shows a considerable warp towards the poles in the outer parts with an absolute difference between inner and outer part (warp amplitude) of up to 10$^\circ$.
	
	Inclination of a polar ring is believed to be a sensitive probe of its internal dynamics. \citet{Sparke1986} studied dynamics and stability of polar rings by approximating PRG with a set of massive concentric wires in a rigid axisymmetric potential. It was found that for systems with $M_*^\text{PS}/M_*^\text{HG}>0.025-0.04$, self-gravity is already strong enough to stabilise the ring. They also showed that upward warps are a natural consequence of a polar ring evolution -- this is how self-gravity counteracts differential precession. \citet{Arnaboldi1994} extended this analysis to a triaxial potential and found that if the potential is highly triaxial then a ring could warp towards the poles instead. \citet{Dubinski1994} used another approach by employing an N-body simulation of a nearly-polar ring in a rigid logarithmic potential. Besides critical mass and warping they also recorded a considerable mass loss of the ring's stellar component into shells. This mass loss also drives the whole ring towards the polar plane as the escaping material carries out $z$-angular momentum. Results of \citet{Christodoulou1992} also show that polar rings are marginally stable over large time scales.
	
	We confirm that PS with relative masses (in gas and stars combined) of at least 0.05 are stable on timescales of $\ge1$Gyr. The classical result of polar ring warping also holds true in TNG50 data even on a quantitative level: warp amplitude of 595100's ring of 10$^\circ$ is consistent with predictions from \citet{Sparke1986} of $\sim8^\circ$. On the other hand, PS found in this study do not show signs of a steady mass loss in contrast to results of \citet{Dubinski1994}. Behaviour of rings' inclination is also different from classical theory -- three rings with the highest age slowly tilt towards the galactic plane. The most probable reason for this disagreement is the frozen potential employed in the mentioned studies. This approach does not allow for any interaction between the ring's particles and the content of the host galaxy and the dark matter halo which is very important. For instance, \citet{Dubinski1995} showed that an inclined stellar disk violently interacts with halo's particles before settling into an equilibrium state. It should also be noted that most PRGs found here have a companion galaxy nearby which in turn perturbes the galactic potential and influences the ring's dynamics. So, to discover what drives the evolution of rings' inclination a detailed study of PRGs dynamics should be conducted which includes influence of both internal and external perturbations.
	
	\subsection{Nuclear activity of PRGs in TNG50}
	Recently, evidence was found for an excess of active nuclei among polar-ring galaxies compared to ordinary galaxies \citep{Smirnov2020}. Among possible explanations of this enhanced nuclear activity is the interaction between PS and the host galaxy or accretion of gas during the polar ring formation. Having access to the full history of galaxies in Illustris, we are able to study the effects of polar ring formation on the nuclear activity of PRGs.
	
	Following \citet{McAlpine2020} and \citet{Kristensen2021}, AGNs are selected based on the Eddington ratio, defined as
	\begin{equation}
		\led\equiv\dot{M}/\dot{M}_\text{Edd},\ \dot{M}_\text{Edd} = \frac{4\pi G M_\text{BH}m_p}{\varepsilon_r \sigma_T c}
	\end{equation}
	where $\dot{M}$ is the actual SMBH mass accretion rate recorded in the IllustrisTNG simulation and $\dot{M}_\text{Edd}$ is the Eddington accretion rate. Galaxies are classified as weak, intermediate and strong AGNs if $0.005\geq\led<0.01$, $0.01\geq\led<0.1$ and $0.1\geq\led$, respectively.
	
\begin{figure}
		\centering
		\includegraphics[width=0.5\textwidth]{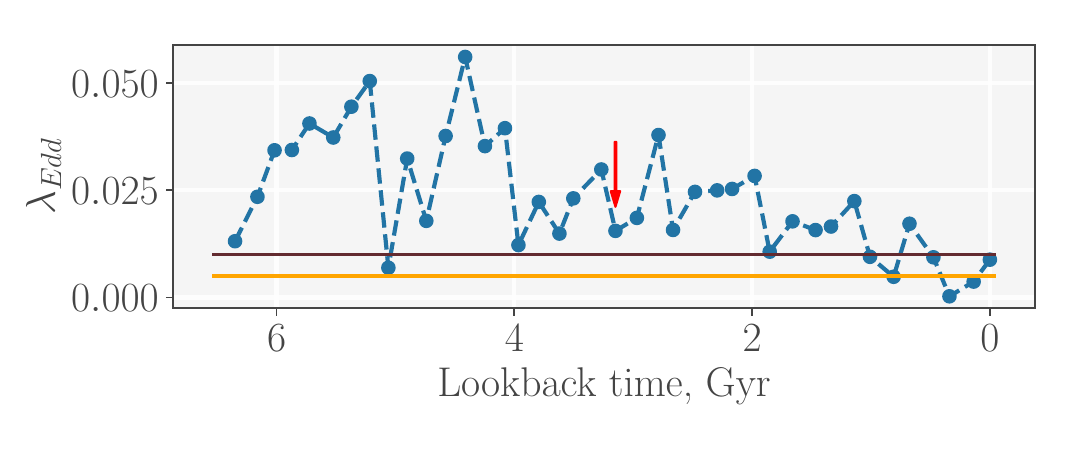}\\
		\includegraphics[width=0.5\textwidth]{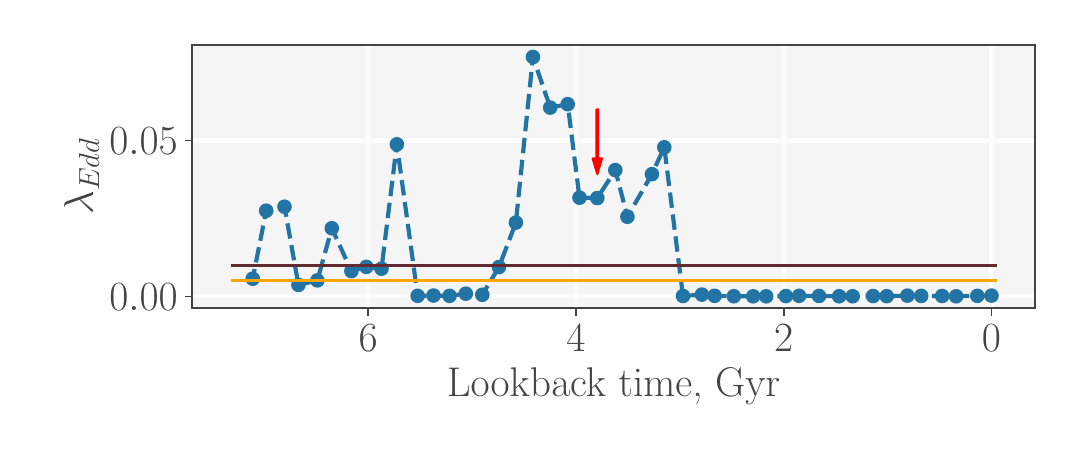}
		\caption{Evolution of Eddington ratio with cosmic time for Subhalo 595100 (\textit{top}) and 501208 (\textit{bottom}). Horizontal lines show the selection thresholds of 0.005 (yellow) and 0.01 (brown). Red arrow marks the beginning of polar ring formation. Analogous figures for other Subhalos are presented in \autoref{fig:AGN_appendix}.}
		\label{fig:AGN}
	\end{figure}

	At $z=0$, only two out of the 6 galaxies in our sample harbour a weak active nucleus with no strong or intermediate AGNs. To study the effects of PS formation, we have plotted the evolution of $\led$ for all the galaxies in our sample, two representative examples of which are set out in \autoref{fig:AGN}. For the galaxy 595100 (top panel) a slight decrease in nuclear activity could be seen, but there is no apparent connection to the ring formation (red arrow). The same is true for the subhalos 428178 and 571908. On the other side, there is a distinct spike in the nuclear activity of 501208 (bottom panel) in the earliest stages of the ring formation when the gas accretion from the companion galaxy begins. This spike lasts for $\sim 1.5$Gyr until the gas disc is destroyed and the central machine is turned off. A similar pattern is observed for the subhalo 483594. 514272 also loses its gas disc, but does not show significant increase in the AGN activity due to gas accretion from the companion.
	
	We can check the validity of these results compared to the observational data from \citet{Smirnov2020}. In that paper we have studied the nuclear activity of polar-ring galaxies from \cite{rm2019} and \cite{moiseev2011} based on data from SDSS and found evidence for possible excess of AGNs among PRGs compared to ``normal" galaxies. It was found that 50\% of best candidates to polar ring galaxies harbour some kind of active nucleus (composite, LINER or Seyfert). If one would draw a random sample of 6 galaxies from these observational data the expected number of AGNs can be easily calculated to be 3.00$\pm$1.16 which is consistent with results obtained in this study. Assuming a naive interpretation of adopted $\led$ cuts as LINER, Seyfert and QSO galaxies, in terms of AGN types TNG50 results are also consistent with observations: 1.85$\pm$1.07 LINERs (weak AGNs), 0.69$\pm$0.75 Seyferts (intermediate AGNs) and 0.12$\pm$0.32 QSOs (strong AGNs). Despite the apparent matching, we recommend caution with these results and strongly argue that they are purely qualitative, due to a small number statistics and intrinsic differences in AGN identification in simulations and observations.
	
	\section{Conclusions}\label{sec:conclusions}
	
	In this paper we present our analysis of a sample of 6 polar-ring galaxies from the TNG50 simulation. TNG50's volume of (51.7\,cMpc)$^3$ is large enough to contain a few PRGs while its resolution allows for a detailed study of the polar structure formation. The conclusions of this work can summarised as follows:
	
	\begin{enumerate}
		\item Searching through the synthetic images of TNG50 galaxies we have found 31 candidates for PRGs. Close inspection of their stellar and gaseous matter distribution revealed that 6 of them are indeed ``true" PRGs. A comparable number of galaxies with exclusively gas PS were found as well as a few related objects with ring structures.
		\item We have successfully performed the kinematic decomposition of the found galaxies using Gaussian Mixture Model clusterization in a ($\jzc,\ \jpc,\ \e$) phase space. Polar structures are characterised by a typical and distinct kinematics: an ordered disk-like rotation in a near polar plane.
		\item Employing supplementary Data Catalogues, it was shown that in terms of their characteristics (integral luminosities, colours, sizes) six found galaxies are very similar to the observed PRGs. Estimates of these properties for PSs and HGs separately are also in general agreement with observations.
		\item We find that a close interaction with a gas-rich companion or a satellite is responsible for the formation of stellar and gaseous polar structures in the studied galaxies. In half of the samples galaxies stellar polar ring was formed due to the star formation in a gas accreted into a polar plane from a companion. In other cases polar structures were formed mainly due to the tidal disruption of the satellite's stellar component.  We report that no galaxies formed in a cold filament accretion or a merger scenario were found. These results support the idea that most polar-ring galaxies are formed in the accretion scenario \citep{bc2003}.
		\item Using the distribution of rings' stellar particles in an edge-on projection we study the evolution of their inclination. A steady increase of the inclination is found for the most massive and long-lived PSs on the contrary to the previous studies. The same approach is used to measure the shape of the rings which reveals consistent warping towards the poles for all galaxies, consistent with classical theory by \citet{Sparke1986}.
		\item Following some observational data on the excess of AGNs among real PRGs, we investigate the evolution of nuclear activity. TNG50 data provide evidence that the formation of polar structures may cause a temporary burst of nuclear activity due to the accretion of gas onto the host galaxy. This implies that the early stages of the PS formation might be correlated with the increase in the nuclear activity of ``young" PRGs. However, on long timescales formation of PS may entirely strip the host galaxy from its gas, turning off the active nucleus.
	\end{enumerate}
	
	\section*{Acknowledgements}
		The authors thank Curtis Struck for the review and useful suggestions which helped improve the manuscript. We would like to acknowledge the work and documentation provided by the IllustrisTNG team that has made this paper possible.

	\section*{Data availability}
	
	The data underlying this article are publicly available at \url{www.tng-project.org/data} \citep{Nelson2019a}. Additional data generated by the analyses in this work will be shared on reasonable request to the corresponding author.
	
	

	\bibliographystyle{mnras}
	\bibliography{art}
	
	\appendix\section{Histories}
	
	\begin{figure*}
		\centering
		\includegraphics[width=0.5\textwidth]{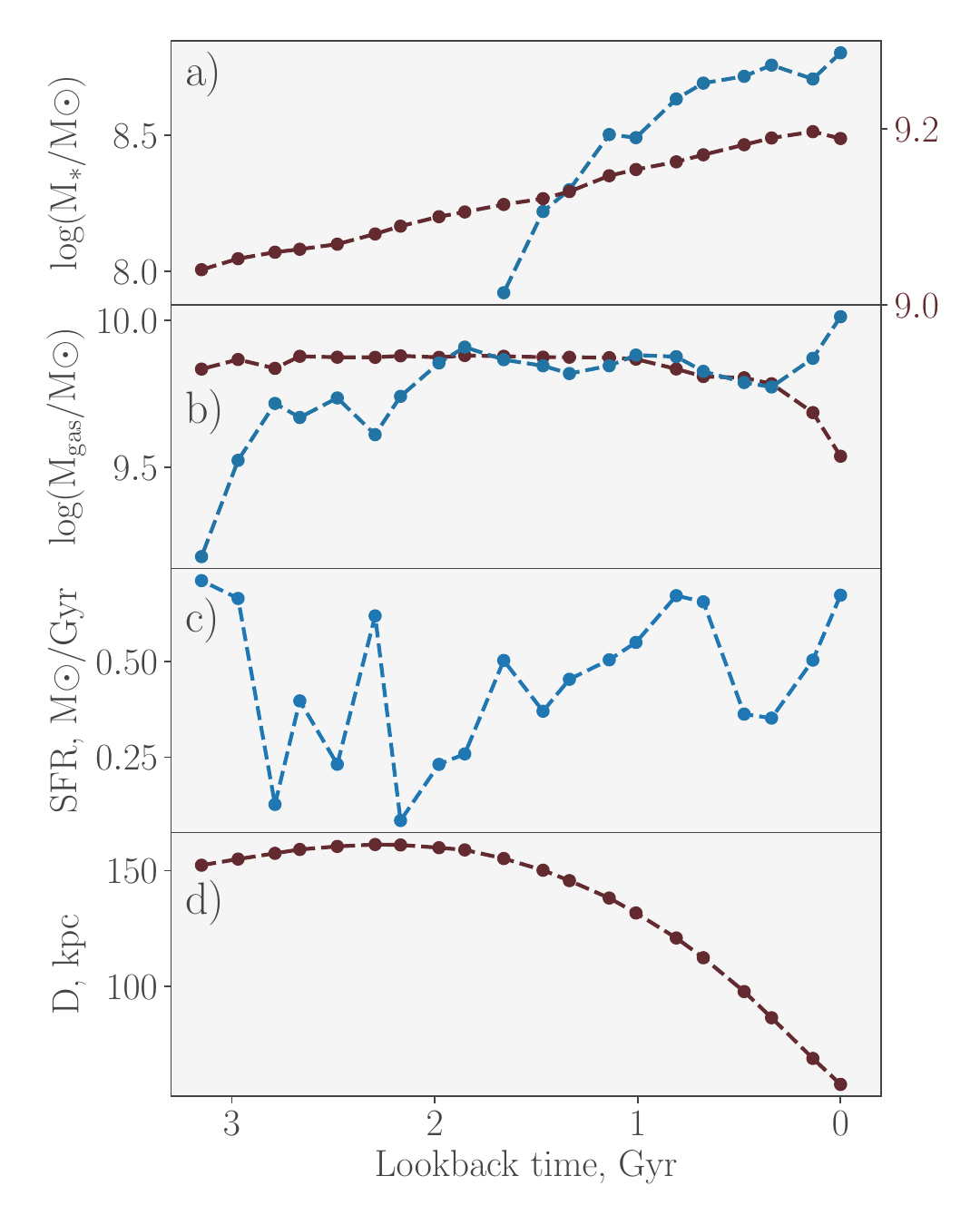}%
		\includegraphics[width=0.5\textwidth]{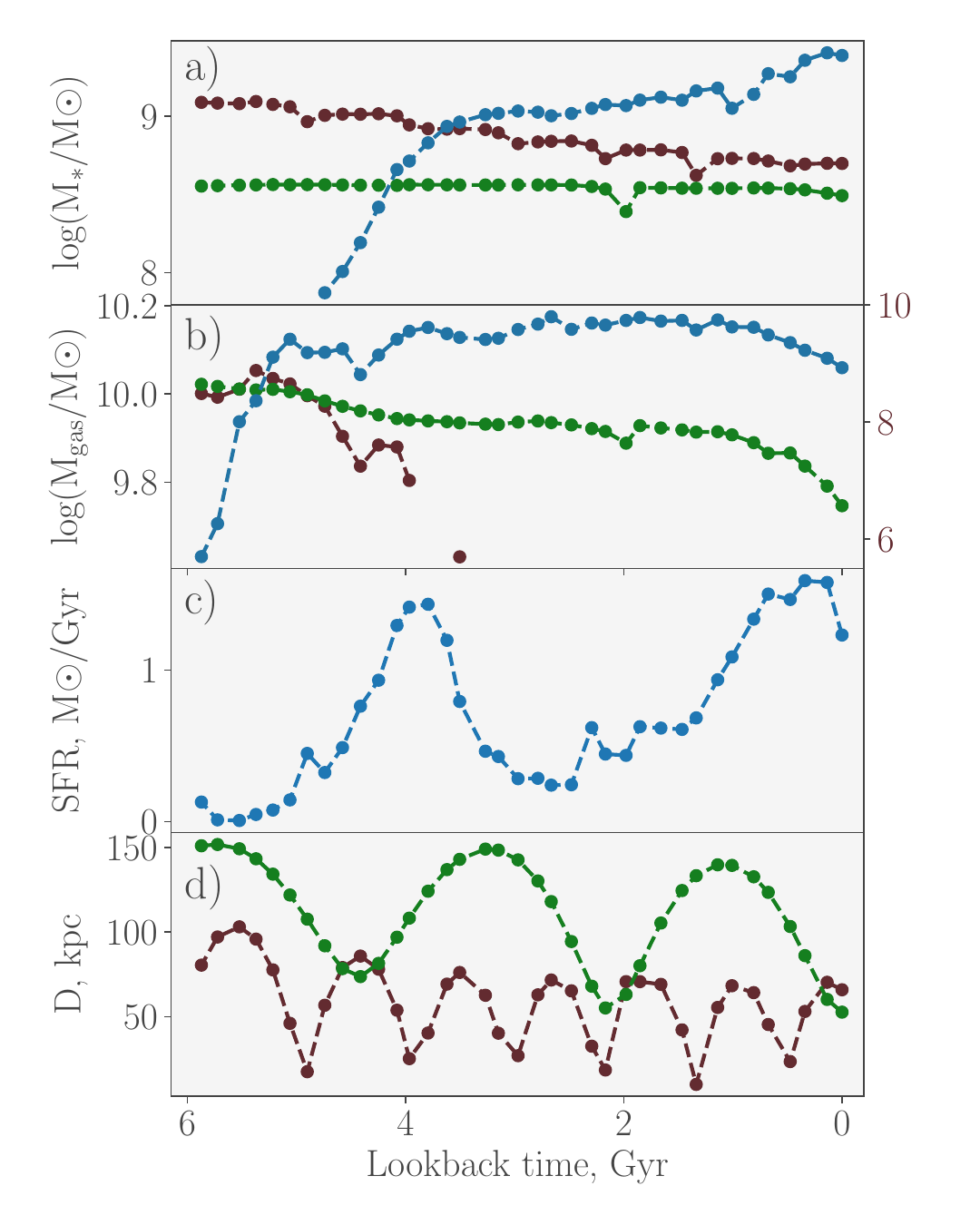}\\
		\includegraphics[width=0.5\textwidth]{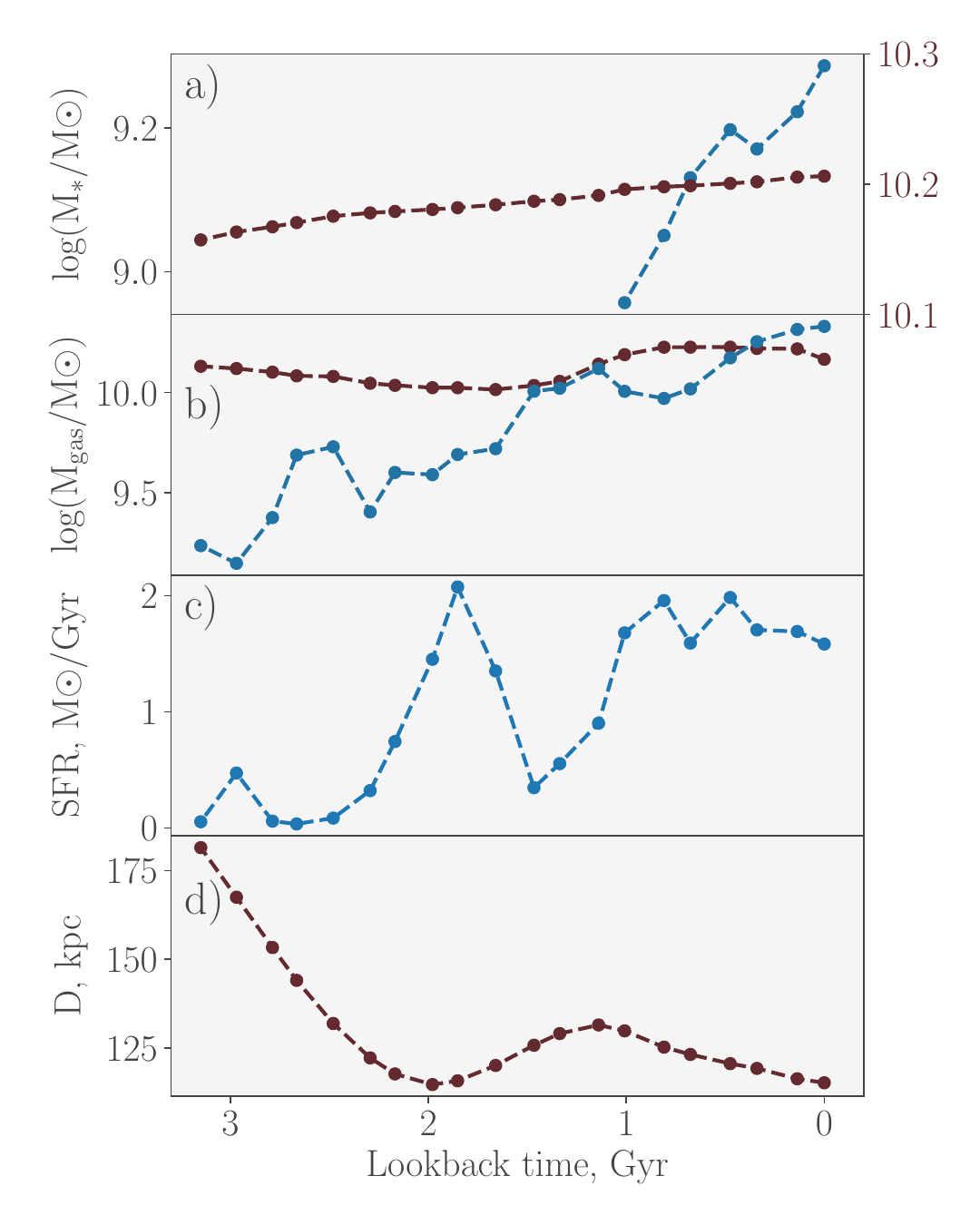}%
		\includegraphics[width=0.5\textwidth]{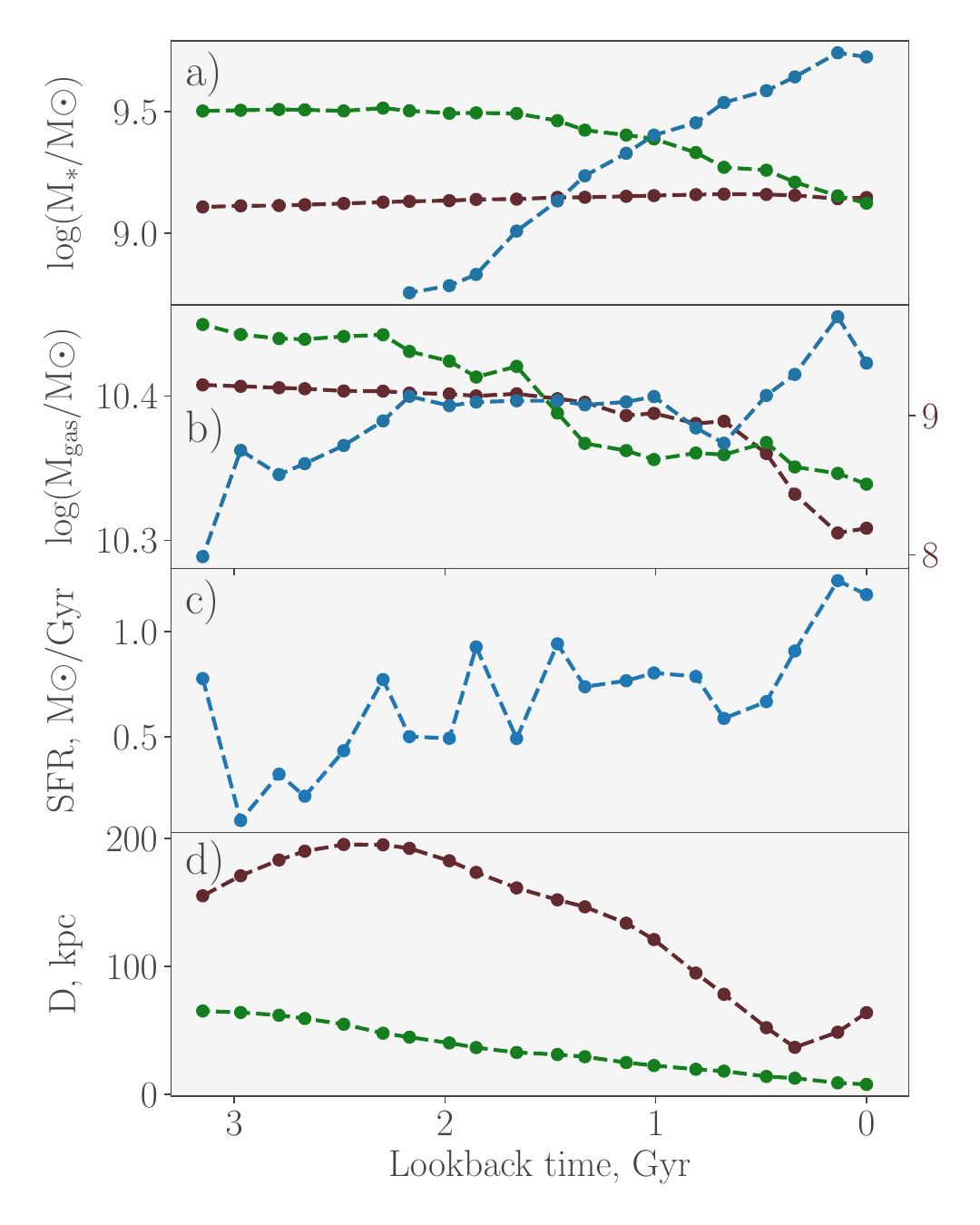}\\
		\caption{Evolution of main parameters of polar structure (blue) and galaxy companion (red and green) of Subhalo 428178 (\textit{top left}), 483594 (\textit{top right}), 514272 (\textit{bottom left}), 571908 (\textit{bottom right}): (a) stellar mass, (b) gas mass, (c) PS star-formation rate and (d) distance between satellite and host galaxy. Ticks on the right show the scale for the properties of satellites in cases where there is a big difference between values for the PS and satellite.}
		\label{fig:history_appendix}
	\end{figure*}
	
	\section{Nuclear activity histories}
	
	\begin{figure*}
		\centering
		\includegraphics[width=0.5\textwidth]{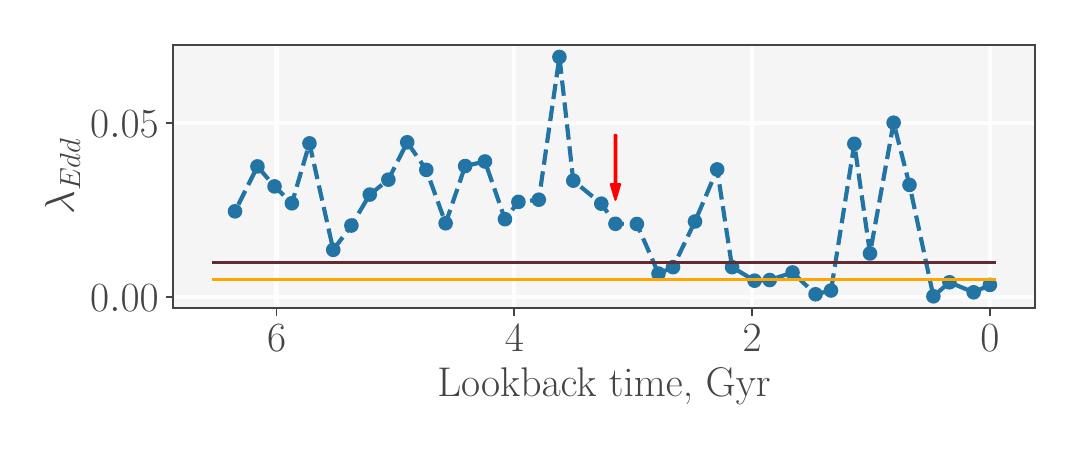}%
		\includegraphics[width=0.5\textwidth]{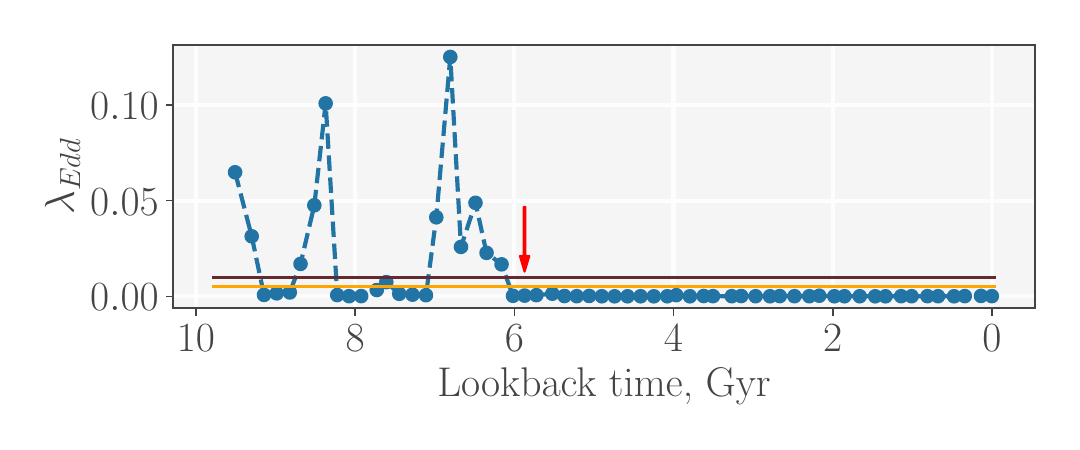}\\
		\includegraphics[width=0.5\textwidth]{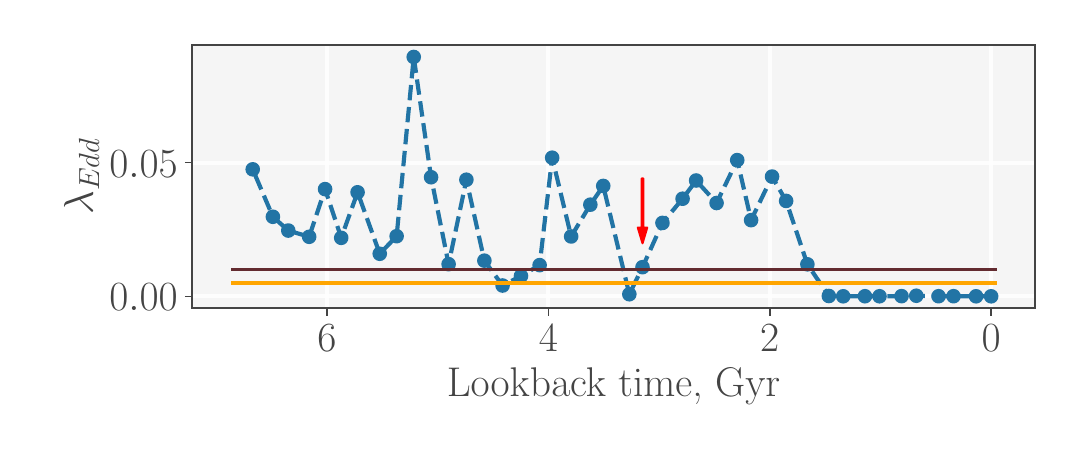}%
		\includegraphics[width=0.5\textwidth]{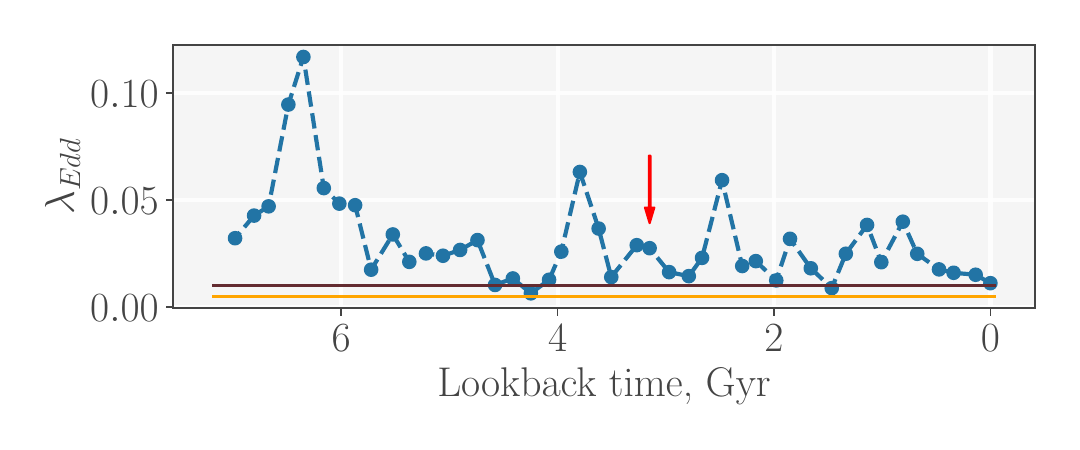}\\
		\caption{Evolution of Eddington ratio with cosmic time for Subhalos 428178 (\textit{top left}), 483594 (\textit{top right}), 514272 (\textit{bottom right}) and 571908 (\textit{bottom right}). Horizontal lines show the selection thresholds of 0.005 (yellow) and 0.01 (brown). Red arrow marks the beginning of polar ring formation.}
		\label{fig:AGN_appendix}
	\end{figure*}	
	
	\section{Inclination}
	
	\begin{figure*}
		\centering
		\includegraphics[width=0.5\textwidth]{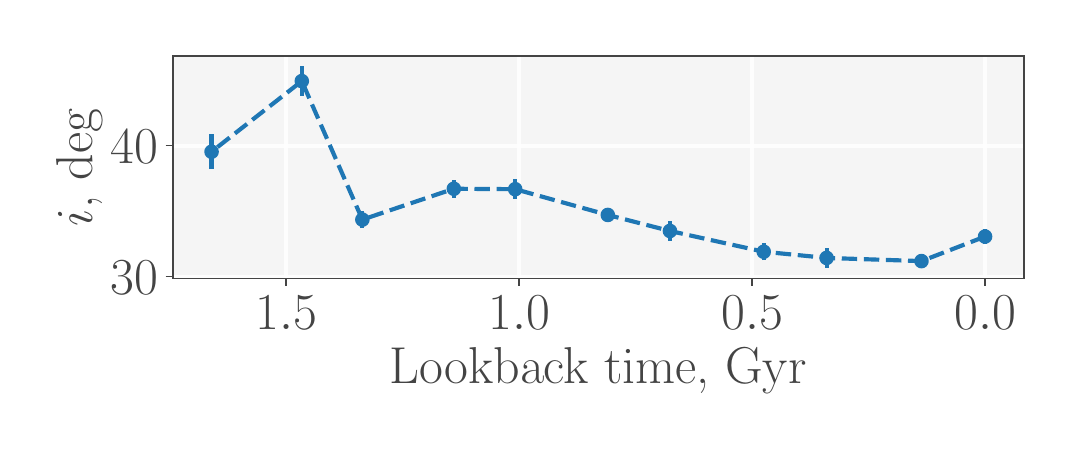}%
		\includegraphics[width=0.5\textwidth]{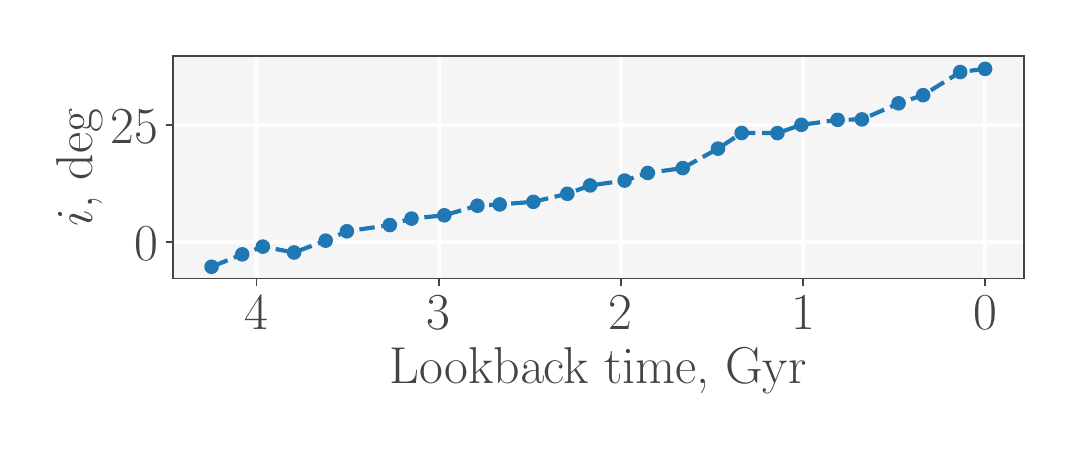}\\
		\includegraphics[width=0.5\textwidth]{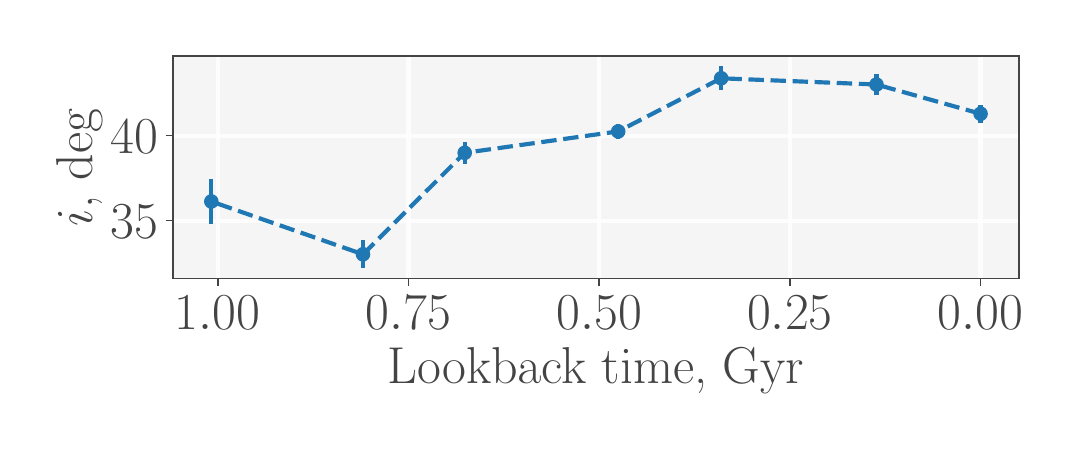}%
		\includegraphics[width=0.5\textwidth]{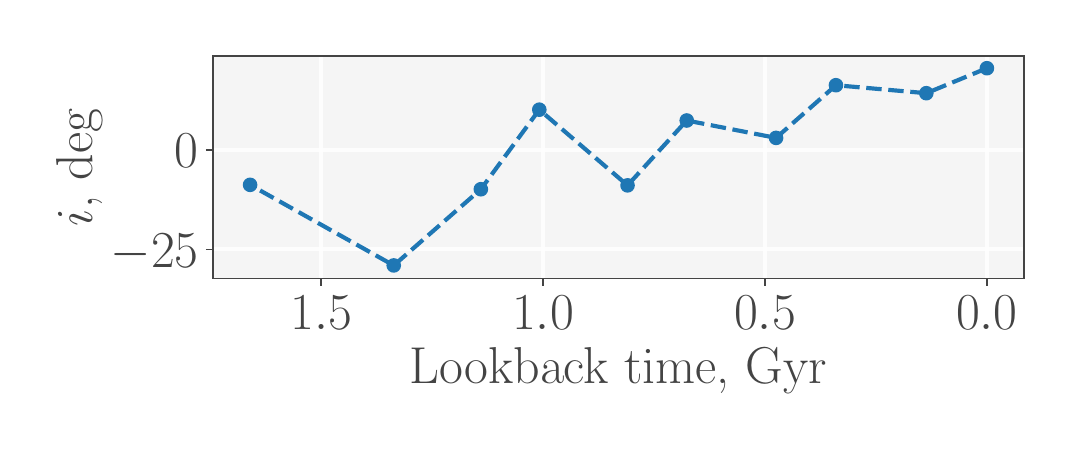}\\
		\caption{Evolution of the angle between the stellar ring and the polar plane with cosmic time for Subhalos 428178 (\textit{top left}), 483594 (\textit{top right}), 514272 (\textit{bottom right}) and 571908 (\textit{bottom right}).}
		\label{fig:inc_appendix}
	\end{figure*}

	\label{lastpage}
\end{document}